\documentclass[twocolumn,showpacs,preprintnumbers,amsmath,amssymb,
showkeys,floatfix,aps]{revtex4}
\usepackage{graphicx}
\usepackage{dcolumn}
\usepackage{bm}
\usepackage{bigints}
\usepackage{natbib}
\usepackage{relsize}
\usepackage{amsmath}
\usepackage{placeins} 

\usepackage{xcolor}
\usepackage{hyperref}

\begin{document}

\title{ Probing CMB Polarization Gaussianity with the Statistics of Unpolarized Points:  Non-Gaussianity of Planck Data and Prospects for Future \textit{B}-Mode Measurements}

\author{Kirill O. Parfenov}
\affiliation{Astro-Space Center of P.N. Lebedev Physical Institute, Profsoyusnaya 84/32, Moscow, 117997}
\author{Dmitry I. Novikov}
\affiliation{Astro-Space Center of P.N. Lebedev Physical Institute, Profsoyusnaya 84/32, Moscow, 117997}
\author{Artem O. Mihalchenko}
\affiliation{Astro-Space Center of P.N. Lebedev Physical Institute, Profsoyusnaya 84/32, Moscow, 117997}

\begin{abstract}

  We present a Gaussianity test of the cosmic microwave background (CMB) polarization by analyzing the statistics
  of unpolarized points in the sky, classified into three distinct types: saddles, comets, and beaks.
  This classification of singular points where both Stokes parameters $Q$ and  $U$ vanish stems from the fact
  that linear polarization is described by a second-rank tensor. By varying the number of spherical
  harmonics included in the polarization maps, one can probe the statistics of these singularities across a range
  of angular scales. Applying this method to Planck data we find clear evidence of non-Gaussianity in both
  E and B modes of polarization. This approach may be especially useful for processing data from current and
  future experiments such as the Simons Observatory (SO). In particular, it can help to assess the Gaussianity
  of a potentially detected B mode
  signal, thereby determining whether it arises from primordial tensor perturbations—as predicted by inflation—or
  from alternative sources such as polarized foregrounds (e.g., thermal dust), E-to-B mode leakage, systematics,
  photon noise or gravitational lensing.  We have made publicly available software that finds
  unpolarized points of all three types on any polarization map in
  Hierarchical Equal Area isoLatitude Pixelation (HEALPix) format with full or incomplete sky coverage
  to enable testing of the observed signal for Gaussianity.
 
\end{abstract}


\maketitle

\section{Introduction}

Inflationary models of
the early Universe predict that the initial perturbations
are Gaussian. Consequently, the observed anisotropy and
polarization of the cosmic microwave background (CMB)
should also follow Gaussian statistics. The
WMAP \citep{2013ApJS..208...20B,2013JCAP...07..018C} and Planck
\citep{2014A&A...571A..15P,2020A&A...641A...4P,2020A&A...641A...7P} missions have provided a
highly accurate measurement of the anisotropy angular power spectrum $C_\ell$ and compelling
evidence that the temperature fluctuations $\Delta T/T$ across the sky generally correspond to
Gaussian distribution.

However, the anomalies of the CMB maps obtained as a result of
multifrequency observations and processing of WMAP and Planck data, both in
temperature \citep{2004PhRvD..70d3515C, 2006MNRAS.367...79C, 2008IJMPD..17..179N,
  2003PhRvD..68l3523T, 2004PhRvL..93v1301S, 2014A&A...571A..15P, 2010AdAst2010E..92C,
  2014PhRvD..90j3510N, 2015arXiv150100282W, 2016A&A...594A..16P, 2016CQGra..33r4001S,
  2021JCAP...03..103C} and in polarization
\citep{2025PhRvD.111f3536N, 2021PhRvD.104b3502K, 2020A&A...641A...7P, 2021JCAP...08..015C,
  2023ApJ...945...79S,2025arXiv250816451B},
indicate some discrepancy between the observed signal phases
and their expected uniform uncorrelated distribution in case of a random Gaussian field.
While various inflation models predict small nonlinear deviations from Gaussianity
\citep{1988PhRvD..38..465O, 1997PhRvD..56..535L}, the observed discrepancy seems largely
attributable to remaining unremoved additional sources of radiation.

Therefore, testing the Gaussianity of observational data serves as a crucial diagnostic -- it verifies whether the signal truly originates from the last scattering surface rather than contaminating foregrounds.
This distinction becomes particularly important for polarization analysis in current and future experiments \citep{2019JCAP...02..056A, 2022ApJ...926...54A, 2023PTEP.2023d2F01L,2025arXiv250816451B}, where detecting the primordial B mode of polarization could provide direct evidence supporting the inflationary hypothesis. Both E and B mode of polarization generated during
recombination through Thomson scattering of anisotropic radiation should exhibit Gaussian statistics. In particular, the Gaussianity of the hypothetically
observed B mode is a criterion that it is caused by tensor perturbations
and represents a signal generated by inflation, provided that it is
uncontaminated by foregrounds, such as  thermal dust emission
\citep{2015PhRvL.114j1301B, 2016A&A...586A.141P} and
undistorted by gravitational lensing \citep{1998PhRvD..58b3003Z, 2006PhR...429....1L}. 

Most of the Gaussianity tests have been developed and conducted in
relation to the study of the scalar anisotropy field on the celestial sphere. Among them are higher-order
correlations 
\citep{1993PhRvL..71.1124L,1994ApJ...427L..71L,2020A&A...641A...9P,2001PhRvD..63f3002K,2022JCAP...03..050G,2020PhRvD.102b3521D,2016PhRvD..94h3503S,2016PhRvD..93l3511M,2016JCAP...05..055B,2016JCAP...03..029B}, Minkowski functionals
\citep{1990ApJ...352....1G,1998MNRAS.297..355S,1998ApJ...507...31N,2014RAA....14..625Z,2015JCAP...02..028G,2016JCAP...07..029S,2017PhLB..771...67C,2019JCAP...01..009J,2021PhRvD.103l3523K,2024MNRAS.527..756C}, percolation \citep{1995ApJ...444L...1N}, kurtosis and skewness \citep{1993ApJ...408...33L}, 
statistics of local extrema (maxima, minima and saddle points)
\citep{2022JCAP...06..006K},
clusterization of peaks \citep{1999IJMPD...8..291N,1998MNRAS.296..693B,2001IJMPD..10..501K,2021MNRAS.503..815V},
the Kullback-Leibler divergence \citep{2015JCAP...06..051B},
 and the
 neural-network approach \citep{2015JCAP...09..064N,2014JCAP...01..018N}.

The Gaussianity
test for polarization is more subtle and complex, since linear polarization is a tensor, which contrasts fundamentally with the scalar field.
 In this work we analyze Planck polarization data by examining the statistics of
 natural singularities -- unpolarized points distributed across the celestial sphere.
 The distinctive patterns formed by the tensor field around zero-polarization points classify these points into three types: saddles, comets, and beaks \citep{2025PhRvD.111f3536N, 1998ApJ...507...31N, 1999IJMPD...8..189D, 2000A&AT...19..213D}. The expected statistics of such singularities for random Gaussian fields was described in
 \citep{1999IJMPD...8..189D}. Any significant deviations from it indicate either non-Gaussianity of the
 initial signal or
 the presence of non-Gaussian foregrounds.

We statistically analyze the distribution of unpolarized points of three different types
 over the sky for Planck polarization data, their relative concentrations, and the distribution of area fractions where the local tensor field satisfies the conditions for each type.
In addition, varying the number of spherical harmonics included in our analysis allows
us to perform the Gaussianity test at different angular scales, yielding a wealth of
scale-dependent information about the origins of possible non-Gaussianities. We show that our approach is extremely sensitive to the presence of foregrounds,
systematics, and distribution of photon noise on polarization maps. Our study shows the presence
of powerful non-Gaussian components in the Planck maps, manifesting themselves at different
angular scales in both the E and B modes.

To find points on the pixelized map where both Stokes parameters $Q$ and $U$ vanish, we use the
algorithm described in \citep{2025PhRvD.111f3536N}. This approach ensures that we can localize and
identify 100\% of the unpolarized points of all three types on a band-limited map (i.e., on a map where the signal does not contain harmonics with $\ell >\ell_{max}$).
To facilitate the convenient and easy
reproduction of our results and the use of our approach in analyzing current and future observations
of CMB polarization, we have made the corresponding software publicly available.
It finds unpolarized points of various types on any  polarization maps with full or incomplete sky coverage
in HEALPix format \citep{2019JOSS....4.1298Z} and also marks map pixels corresponding to three
different types. The Fortran program is included in
\footnote{\url{https://github.com/koparfenov169-lab/CMB-polarization-Gaussianity-test}}
with detailed instruction. 

To study Gaussianity at different scales, we have to vary the number of harmonics 
used in the map. Therefore, for fast simulation of polarization maps, we modified the approach to
spherical harmonic transform used in HEALPix \citep{2005ApJ...622..759G}. In order to speed the
process up we used a recurrence relation for the associated Legendre polynomials on $m$ with fixed
$\ell$ instead of the recurrence relation on $\ell$ with fixed $m$.  If we want to study a map in which
only harmonics with a single fixed $\ell$ are present (i.e., those that are responsible for the angular scale
corresponding to $\ell$), then in this case we can simulate such a map in $O(\ell^2 \log_{{}_2}\ell)$ operations
instead of $O(\ell^3)$, as in HEALPix.

The outline of this paper is as follows: In Sec. II we review the classification of the
singular points of the CMB polarization and their statistical properties
for Gaussian random fields. We also generalize this approach to E and B polarizations taking into
account spherical geometry. In Sec. III we present the results of a Gaussianity test for
Planck CMB polarization data \citep{2020A&A...641A...4P},
analyzing the distribution of different types of points over the sky,
as well as the distribution of the corresponding areas. By changing the resolution,
i.e., by varying the maximum number of harmonics used,
we carry out this analysis at all angular scales. In the same Section we provide,
as an example, the same analysis for the E polarization on a sphere, caused by the
highly non-Gaussian scalar field of the Earth's surface.
Finally, in Sec. IV we summarize our results and make brief conclusions. Separately in the
Appendix we describe our approach to fast spherical harmonics transform and the method for its
stabilization.

\section{Singular (unpolarized) points of CMB polarization field}

 \subsection{Classification of unpolarized points in the sky}

 \begin{figure}[!htbp]
   \includegraphics[width=1.0\columnwidth]{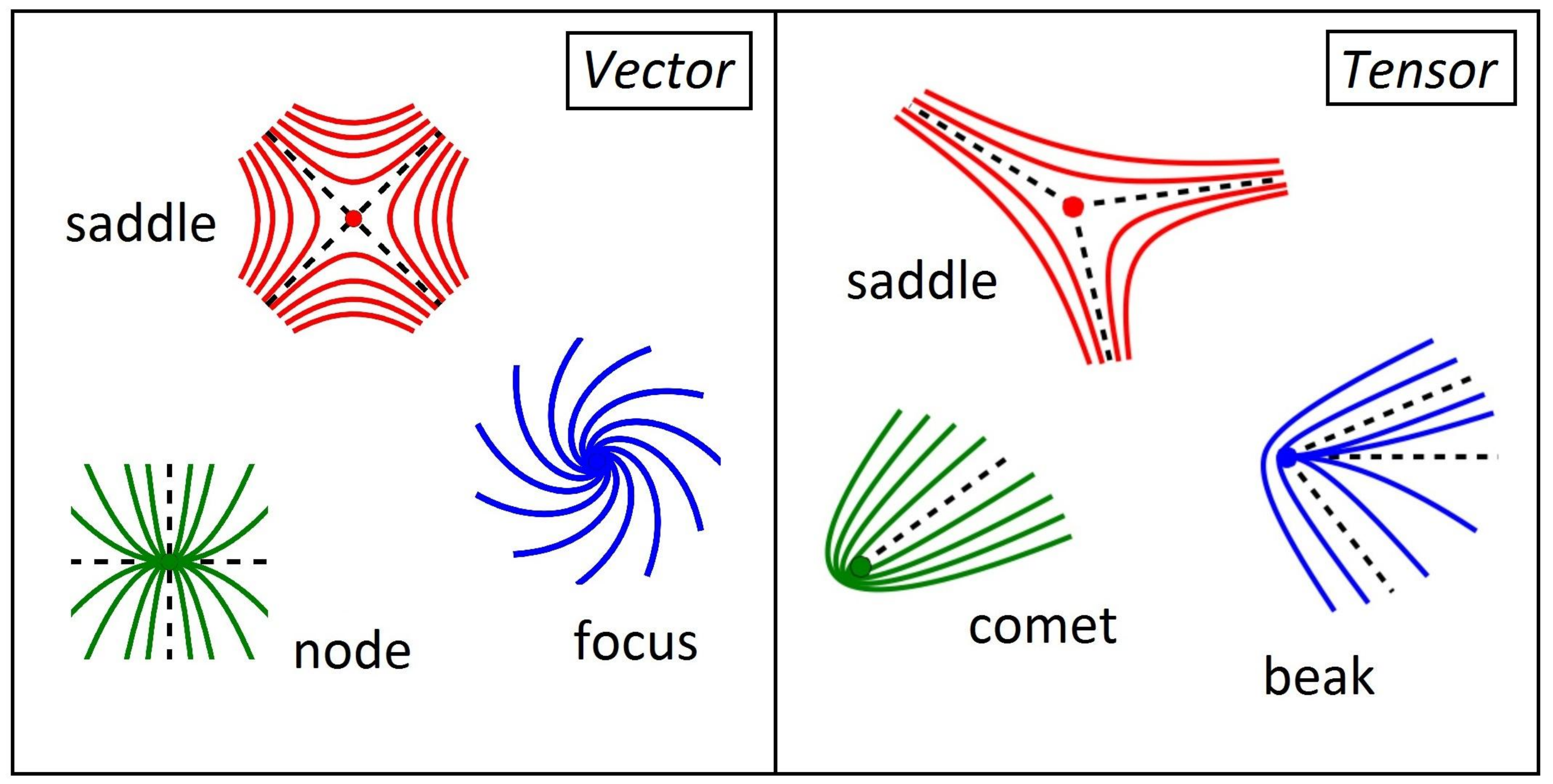}
   \caption{Singularities in vector and tensor fields.
     {\it Left panel:} two-dimensional velocity vector field ${\bf V}$.
     Continuous lines show the orientation of vectors in the vicinities of points where
     ${\bf V}=0$ (vector directions are not indicated). Foci cannot exist in a
     gradient field $\nabla\times{\bf V}=0$, and nodes do not exist in a vortex field $\nabla\cdot{\bf V}=0$.
     {\it Right panel:} two-dimensional tensor field.
     Lines show the orientation of polarization in the vicinities of points with zero polarization: $Q=U=0$.
     All three types of points can exist in both E and B modes.} 
\end{figure}
 
Linear polarization is a second rank traceless tensor and is described by
two Stokes parameters $Q$ and $U$. It is convenient to visualize this
tensor in terms of the degree of polarization
$P$ and its orientation $\phi$ relative to a local coordinate
system $(x,y)$, which is perpendicular to the radiation
propagation,
\begin{equation}
    \begin{array}{l}
     P^2=Q^2+U^2,\hspace{1cm}\tan(2\phi)=U/Q. 
     \end{array} 
\end{equation}
By choosing this coordinate system on the celestial sphere such
that $x$ is directed along the
meridian from south to north, and $y$ along the great circle from east
to west, one can consider an extremely small area of the sky in the
geometrically flat limit. Thus we are dealing with a two-dimensional
tensor field, where $Q=Q(x,y)$, $U=U(x,y)$.
We are interested in singular points in the sky where
the polarization vanishes: $P(x_0,y_0)=0$.
Since both Stokes parameters are zero at such a point, their values in its
vicinity can be written as,
\begin{equation}
  \begin{array}{l}
    \begin{pmatrix}Q\\U\end{pmatrix}=
    \begin{pmatrix}Q_x&Q_y\\U_x&U_y\end{pmatrix}\cdot
     \begin{pmatrix}\Delta x\\ \Delta y\end{pmatrix},
   \end{array}
\end{equation}
where $\Delta x=x-x_0$, $\Delta y=y-y_0$ and $Q_x$, $U_x$, $Q_y$, $U_y$ are
derivatives of Stokes parameters over $x$ and $y$ at point $(x_0,y_0)$. Depending on the values of these
derivatives, the orientation of the polarization field
around such a point creates a characteristic pattern that determines the
type of singularity.
According to \citep{1999IJMPD...8..189D} the classification of singular points depends on the
signs of the determinant $D$ and discriminant $\Delta$,
\begin{equation}
   \begin{array}{l}
     \vspace{0.3cm}
     D=Q_xU_y-U_xQ_y,\\
     \vspace{0.2cm}
     \Delta= 4(U_x+2Q_y)^3U_x+\\
     \vspace{0.2cm}
    (U_x+2Q_y)^2(2Q_x-U_y)^2-
     4U_y(2Q_x-U_y)^3-\\
     18U_y(U_x-2Q_y)(2Q_x-U_y)U_x-27U_x^2U_y^2.
   \end{array}
\end{equation}
The expressions for the determinant and discriminant come from solving the corresponding cubic equation.
For more details, see \citep{2000A&AT...19..213D}.
There are three different types of points.
If $D<0$ then such a point is a
 ``saddle". For the case when $D>0$ there are two possibilities.
 If $\Delta<0$ then this point is a ``comet". Otherwise,
 if $\Delta>0$, then it is a ``beak" (see Fig 1).

 The derivatives with respect to $x$, $y$ should be considered as covariant
 and in spherical coordinates $(\theta,\varphi)$ look as follows:
\begin{equation}
  \begin{array}{l}
    \vspace{0.2cm}
    Q_x=Q_{\theta},\hspace{0.5cm}U_x=U_{\theta},\\
    \vspace{0.2cm}
    Q_y=\frac{1}{\sin\theta}Q_{\varphi}-
    2\frac{\cos\theta}{\sin\theta}U,\\
    \vspace{0.4cm}
    U_y=\frac{1}{\sin\theta}U_{\varphi}+
    \frac{2\cos\theta}{\sin\theta}Q.\\
    \vspace{0.2cm}
    \end{array}
\end{equation}
Stokes parameters can be completely described in terms of
the scalar $E(x,y)$ and the pseudoscalar $B(x,y)$ fields \citep{2000A&AT...19..213D}:
\begin{equation}
  \begin{array}{l}
    \vspace{0.2cm}
    Q=Q^E+Q^B,\hspace{0.5cm}U=U^E+U^B,\\
    \vspace{0.2cm}
    Q^E=E_{xx}-E_{yy},\hspace{0.5cm}U^E=2E_{xy},\\
    Q^B=-2B_{xy},\hspace{0.5cm}U^B=B_{xx}-B_{yy}.\\    
    \end{array}
  \end{equation}
Here two-dimensional functions $E$ and $B$ correspond to the ``electric" and 
``magnetic" modes of polarization. In the $\theta,\varphi$ coordinates,
the operators in Eq. (5) have the following form \citep{2025PhRvD.111f3536N}:
\begin{equation}
  \begin{array}{l}
    \vspace{0.2cm}
    Q^E=E_{\theta\theta}-\frac{\cos\theta}{\sin\theta}E_{\theta}-
    \frac{1}{\sin^2\theta}E_{\varphi\varphi},\\
    \vspace{0.2cm}
    U^E=\frac{2}{\sin\theta}\left(E_{\theta\varphi}-\frac{\cos\theta}
    {\sin\theta}E_{\varphi}\right),\\
    \vspace{0.2cm}
     Q^B=-\frac{2}{\sin\theta}\left(B_{\theta\varphi}-\frac{\cos\theta}
    {\sin\theta}B_{\varphi}\right),\\
    U^B=B_{\theta\theta}-\frac{\cos\theta}{\sin\theta}B_{\theta}-
    \frac{1}{\sin^2\theta}B_{\varphi\varphi}.\\
    \end{array}
\end{equation}

All three types of points
 can be present both in the polarization field generated by the scalar
 (E mode) and in the polarization caused by the pseudoscalar (B mode).
 This is a natural property of a two-dimensional tensor field. Note that in
 a two-dimensional vector field (for example, a velocity field) there can
 also be singular points where the velocities vanish. Their classification
 is well known \citep{strogatz:2000}. They are also divided into three types: nodes, saddles,
 and foci. However, in a gradient vector field (E mode analogy) there cannot
 be singularities of the focus type.

\subsection{Statistics of three types of unpolarized points
   in a random Gaussian field}

\begin{figure}[!htbp]
  \includegraphics[width=0.49\columnwidth]{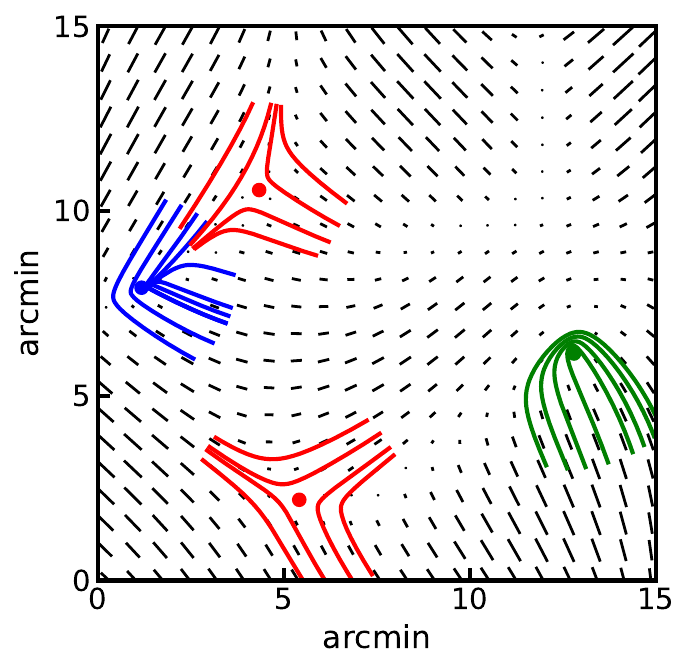}
  \includegraphics[width=0.49\columnwidth]{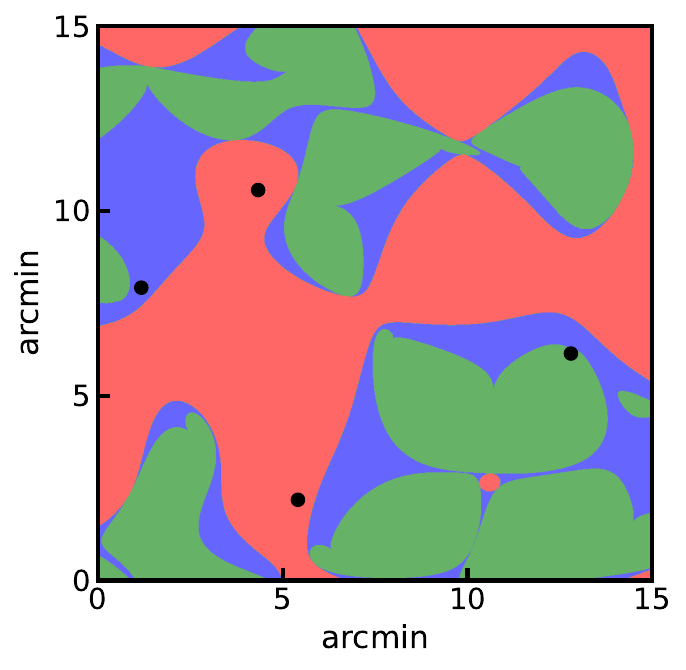}
   \caption{Polarization map of a small $15'\times 15'$ part of the sky with unpolarized points.
     {\it Left panel:}  the orientation of each segment corresponds to the orientation of linear polarization,
     and the lengths of the segments are proportional to the polarization level $P$.
     Singularity points: saddles are marked in red, comets in green, and beaks in blue, respectively.
     {\it Right panel:} the same part of the sky. Areas satisfying the conditions for saddles, comets
     and beaks are colored red, green, and blue, respectively. The same singularity points as in the left
     panel are marked in black.
   } 
\end{figure}

\begin{figure*}[!htbp]
   \includegraphics[width=0.49\textwidth]{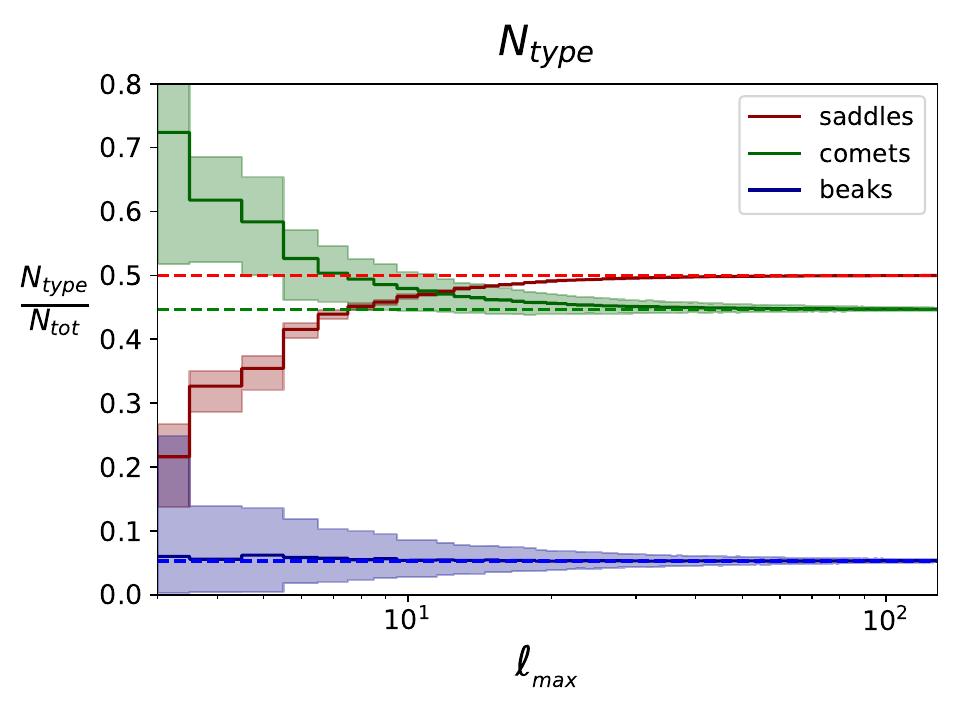}   
   \includegraphics[width=0.49\textwidth]{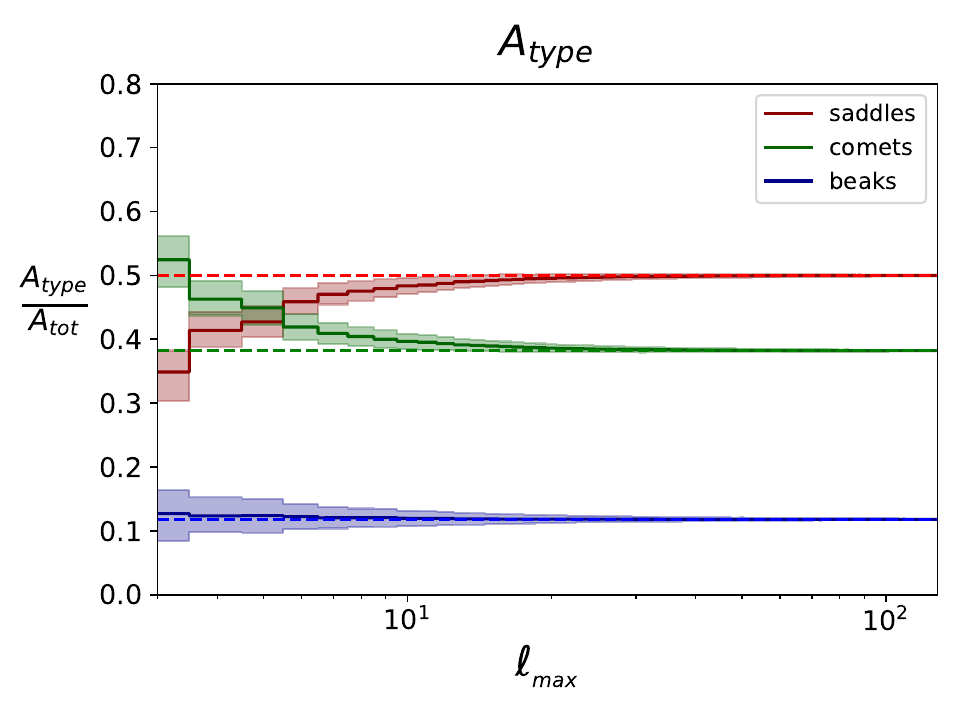}
   \caption{Expected ratios $\frac{N_{type}}{N_{tot}}$ for saddles, comets, and beaks
     ({\it left panel}) and expected fractions of area corresponding to saddle, comet, and beak conditions
     ({\it right panel}) for a random Gaussian field. The results are obtained for the CMB E polarization spectrum
    according to SMICA 2018 Planck data. Solid lines
     show the mean expected values of the $\frac{N_{type}}{N_{tot}}(\ell_{max})$, 
     $\frac{A_{type}}{A_{tot}}(\ell_{max})$ functions. Dashed lines correspond to the asymptotic values
     in a flat limit. The shaded areas in both panels show the standard
     $\sigma_{N_{type}}^\pm$ and $\sigma_{A_{type}}^\pm$ deviations for the all-sky statistics.}
\end{figure*}

The random scalar $E$ and pseudoscalar $B$ fields of Eq. (5) determine the
statistics of polarization and, as a consequence, the distribution of
different types of unpolarized points on the celestial sphere
including their concentration per unit area. 
The functions E and B can be decomposed into spherical harmonics,
\begin{equation}
  \begin{array}{l}
    E,B=\sum\limits_{\ell=\ell_{min}}^{\ell_{{}_{max}}}\sum\limits_{m=-\ell}^{\ell}
    a_{\ell m}^{E,B}\cdot Y_{\ell m}(\boldsymbol{\eta}),\hspace{0.2cm}\boldsymbol{\eta}=(\theta,\varphi),
    \end{array}
\end{equation}
where $\ell_{min}$ and $\ell_{max}$ limit the total number of harmonics represented on the map:
$\ell_{min}\le\ell\le\ell_{max}$. The expansion
coefficients $a_{\ell m}^{E,B}$ are related to the power spectrum
$C_\ell=C_\ell^E+C_\ell^B$ as follows:
\begin{equation}
    \begin{array}{l}
   C_\ell=\frac{1}{2\ell+1}\sum\limits_{m=-\ell}^\ell\left[
     a_{\ell m}^{E}\cdot a_{\ell m}^{E\mathlarger{\mathlarger{\mathlarger{\ast}}}}+
     a_{\ell m}^{B}\cdot a_{\ell m}^{B\mathlarger{\mathlarger{\mathlarger{\ast}}}}\right],
   \end{array}
\end{equation}
where the asterisk denotes the complex conjugate.
We introduce the following spectral parameters:
\begin{equation}
  \begin{array}{l}
    \vspace{0.3cm}
    \sigma_0^2=\sum\limits_{\ell=\ell_{min}}^{\ell_{max}}(2\ell+1)
    \frac{(\ell+2)!}{(\ell-2)!}(C_\ell^E+C_\ell^B),\\
    \sigma_1^2=\sum\limits_{\ell=\ell_{min}}^{\ell_{max}}(2\ell+1)
    \frac{(\ell+3)!}{(\ell-3)!}
    \left(C_\ell^E+C_\ell^B\right).
   \end{array}
\end{equation}
In terms of these parameters, it is convenient to use dimensionless quantities,
\begin{equation}
   \begin{array}{l}
     \vspace{0.2cm}
     q=\frac{Q}{\sigma_0},\hspace{0.2cm}u=\frac{U}{\sigma_0},\hspace{0.2cm}
     p=\frac{P}{\sigma_0}\\
     \vspace{0.2cm}
     q_x=\frac{Q_x}{\sigma_1},\hspace{0.2cm}q_y=\frac{Q_y}{\sigma_1},
     \hspace{0.2cm}u_x=\frac{U_x}{\sigma_1},
     \hspace{0.2cm}u_y=\frac{U_y}{\sigma_1},\\
     d=\frac{D}{\sigma_1^2},\hspace{0.5cm}\delta=\frac{\Delta}{\sigma_1^4}.
   \end{array}
 \end{equation}
If the polarization is generated by random independent Gaussian fields
$E$ and $B$, then the random values $q,u,q_x,q_y,u_x,u_y$ are
uncorrelated with each other, and have zero means and variances, 
\begin{equation}
   \begin{array}{l}
     \vspace{0.2cm}
     \langle q^2\rangle=\langle u^2\rangle=1,\\
     \langle q_x^2\rangle=\langle q_y^2\rangle=
     \langle u_x^2\rangle=\langle u_y^2\rangle=\frac{1}{2}
   \end{array}
\end{equation}
According to \citep{1998ApJ...507...31N} the average number density of unpolarized points
is calculated as follows:
\begin{equation}
   \begin{array}{l}
     \vspace{0.2cm}
     \langle n_{type}\rangle=\frac{1}{2\pi^3}\frac{\sigma_1^2}{\sigma_0^2}
     \int\limits_{\Omega_{type}} e^{-g}\mid d\mid dq_xdq_ydu_xdu_y,\\
      \vspace{0.2cm}
      g=q_x^2+q_y^2+u_x^2+u_y^2,\\
   \end{array}
\end{equation}
where the index ``$type$" denotes the type of singularity, and $\Omega_{type}$
defines the boundaries of integration:\\
i. Total number density of all unpolarized points $n_{type}=n_{tot}$:
$d$ and $\delta$ are arbitrary;\\
ii. ``saddles" $n_{type}=n_{s}$: $d<0$;\\
iii. ``comets" $n_{type}=n_{c}$: $d>0$, $\delta<0$;\\
iv. ``beaks" $n_{type}=n_{b}$: $d>0$, $\delta>0$.

Integration of the right-hand side of Eq. (12) can be easily accomplished in two steps. Finding the
expected total number density of unpolarized points (when $d$ and $\delta$ are arbitrary) can be
done analytically (see \citep{2025PhRvD.111f3536N}). The remainder of the problem is to find the proportions of points of different types
$\langle N_{type}\rangle/\langle N_{tot}\rangle$. It is more convenient to solve this problem
without directly calculating the integral, but by repeatedly generating four independent
Gaussian random numbers  $q_x,q_y,u_x,u_y$  with zero mean and mean squares equal to $1/2$
[see Eq. (11)].
For each generated quadruple of numbers we must find the value $\omega=\mid d\mid$ and,
depending on the signs of $\delta$ and $d$, assign this quadruple to a point of a specific type:
(saddle, comet, beak) with a weight equal
to $\omega$. A large number of simulated random quadruples of numbers, treated as unpolarized
points, yields proportions of points of three different types with the required accuracy.

Thus, after calculations we obtain that for a part of the sky with area A
the expected numbers of points of different types $N_{type}=A\cdot n_{type}$ are as
follows:
\begin{equation}
  \begin{array}{l}
    \vspace{0.2cm}
     \langle N_{tot}\rangle=\frac{A}{2\pi^2r_c^2},\hspace{0.5cm}
     \langle N_{s}\rangle=0.5\cdot\langle N_{tot}\rangle,\\
     \langle N_{c}\rangle=0.447\cdot\langle N_{tot}\rangle,\hspace{0.5cm}
     \langle N_{b}\rangle=0.053\cdot\langle N_{tot}\rangle,     
\end{array}
\end{equation}
where $r_c=\frac{\sigma_0}{\sigma_1}$ is the correlation scale and
$A=4\pi$ for the whole sky.

Another important statistical characteristic is the distribution of area fractions where the conditions for
derivatives satisfy saddles, comets, and beaks, but the
values of the Stokes parameters themselves can be arbitrary. The expression for the average fractions of such areas is as follows:
\begin{equation}
   \begin{array}{l}
     \langle A_{type}\rangle=\frac{A}{\pi^2}\int\limits_{\Omega_{type}}
     e^{-g}dq_xdq_ydu_xdu_y.
   \end{array}
\end{equation}
The integration can be done similarly to that described above for Eq. (12), but in this case each
quadruple of random numbers is considered as a pixel and weighted equally with a weight of $\omega=1$,
regardless of the value of $\mid d\mid$. It gives the following results:
\begin{equation}
  \begin{array}{l}
    \vspace{0.2cm}
     A_{tot}=A,\hspace{0.5cm}\langle A_s\rangle=0.5\cdot A_{tot},\\
     \langle A_c\rangle=0.382\cdot A_{tot},\hspace{0.5cm}
     \langle A_b\rangle=0.118\cdot A_{tot}.
   \end{array}
\end{equation}
Comparing Eqs. (13), (15), one can conclude that the unpolarized points are not uniformly distributed across the sky.
Indeed,
\begin{equation}
  \begin{array}{l}
    \vspace{0.2cm}
    \langle N_s\rangle/\langle A_s\rangle=\langle n_{tot}\rangle,\hspace{0.3cm}
    \langle N_c\rangle/\langle A_c\rangle=1.167\cdot\langle n_{tot}\rangle,\\
    \langle N_b\rangle/\langle A_b\rangle=0.448\cdot\langle n_{tot}\rangle, \hspace{0.2cm}
    \langle n_{tot}\rangle=\frac{1}{2\pi^2r_c^2}.
   \end{array}
\end{equation}
With a uniform distribution, the number of points of different types per unit of their corresponding area would be
constant: $\langle N_{type}\rangle/\langle A_{type}\rangle=\langle n_{tot}\rangle$. Nevertheless, the number of points of different types
per corresponding area varies depending on the type of points. The area of
beaks is less populated than the areas of comets and saddles. The comets area is the most
populated and the concentration of points on it exceeds the average concentration
$\langle n_{tot}\rangle$.

Note that the results of Eqs. (13), (15), (16) do not depend on the power spectrum
(except for the total number of unpolarized points) and are exclusively a consequence
of the Gaussian nature of the signal. However, these results are obtained for a flat geometry,
i.e., when the correlation size is much smaller than the radius of the sphere.
This condition is equivalent to $\ell_{max}\gg 1$ .
For small $\ell_{max}$ the values of the concentration of points of different
types and the fraction of the corresponding areas depend on the spectrum $C_\ell$ and must be
obtained numerically for a Gaussian random field with a given $C_\ell$
and $\ell_{min}\le\ell\le\ell_{max}$.

The harmonic range $(\ell_{min},\ell_{max})$ can be varied to study the Gaussianity of the signal at different
angular scales.  Using the algorithm described in \citep{2025PhRvD.111f3536N},
one can find unpolarized points of three different types on the simulated Gaussian maps with
a given power spectrum $C_{\ell}$. In our studies of Planck maps we restrict ourselves to a fixed
lower limit of $\ell_{min}=2$.
Thus, by successively increasing the number of harmonics used in the map,
we construct functions $N_s(\ell_{max})$, $N_c(\ell_{max})$, $N_b(\ell_{max})$ for the
saddles, comets, and beaks respectively. In this way we find the dependence of the total number
of points of a certain type on the number of harmonics used.
The larger number of harmonics that makes up the map, the smaller the correlation radius becomes
and the more unpolarized points appear. Similarly, we find the result for the
areas corresponding to saddles, comets, and beaks:
$A_s(\ell_{max})$, $A_c(\ell_{max})$, $A_b(\ell_{max})$.
For large $\ell_{max}$  in the case of a Gaussian random field these functions tend to the asymptotic
values of equations of Eqs. (13), (15) (see Fig. 3). Using a large number of simulations for each $\ell_{max}$,
one can find the average expected values for  $\langle N_{type}(\ell_{max})\rangle$, 
$\langle A_{type}(\ell_{max})\rangle$, and standard deviations from the average,
\begin{equation}
  \begin{array}{l}
    \vspace{0.2cm}
    \sigma_{N_{type}}^{\pm}=\sqrt{\langle \left(N_{type}^{\pm}\right)^2\rangle-\langle N_{type}\rangle^2},\\
    \sigma_{A_{type}}^{\pm}=\sqrt{\langle \left(A_{type}^{\pm}\right)^2\rangle-\langle A_{type}\rangle^2},
    \hspace{0.2cm}type=s,c,b,
   \end{array}
\end{equation}
where + and - denote positive and negative deviations from the average. Note that
$\sigma_{N_{type}}^{+}>\sigma_{N_{type}}^-$ and  $\sigma_{A_{type}}^{+}>\sigma_{A_{type}}^-$.
If $\ell_{max}\gg 1$ then  $\sigma_{N_{type}}^{+}\rightarrow\sigma_{N_{type}}^-$ and
$\sigma_{A_{type}}^{+}\rightarrow\sigma_{A_{type}}^-$. 

The results described above should be calculated for a Gaussian field with the spectrum
$C_\ell$ observed in the experiment. Their comparison with the statistics of the observed
polarization makes it possible to detect non-Gaussianity at various scales. In Fig. 3 we show the numerical
results of our calculation for the CMB Planck SMICA polarization spectrum.
\begin{figure*}[!htbp]
  \includegraphics[width=0.49\textwidth]{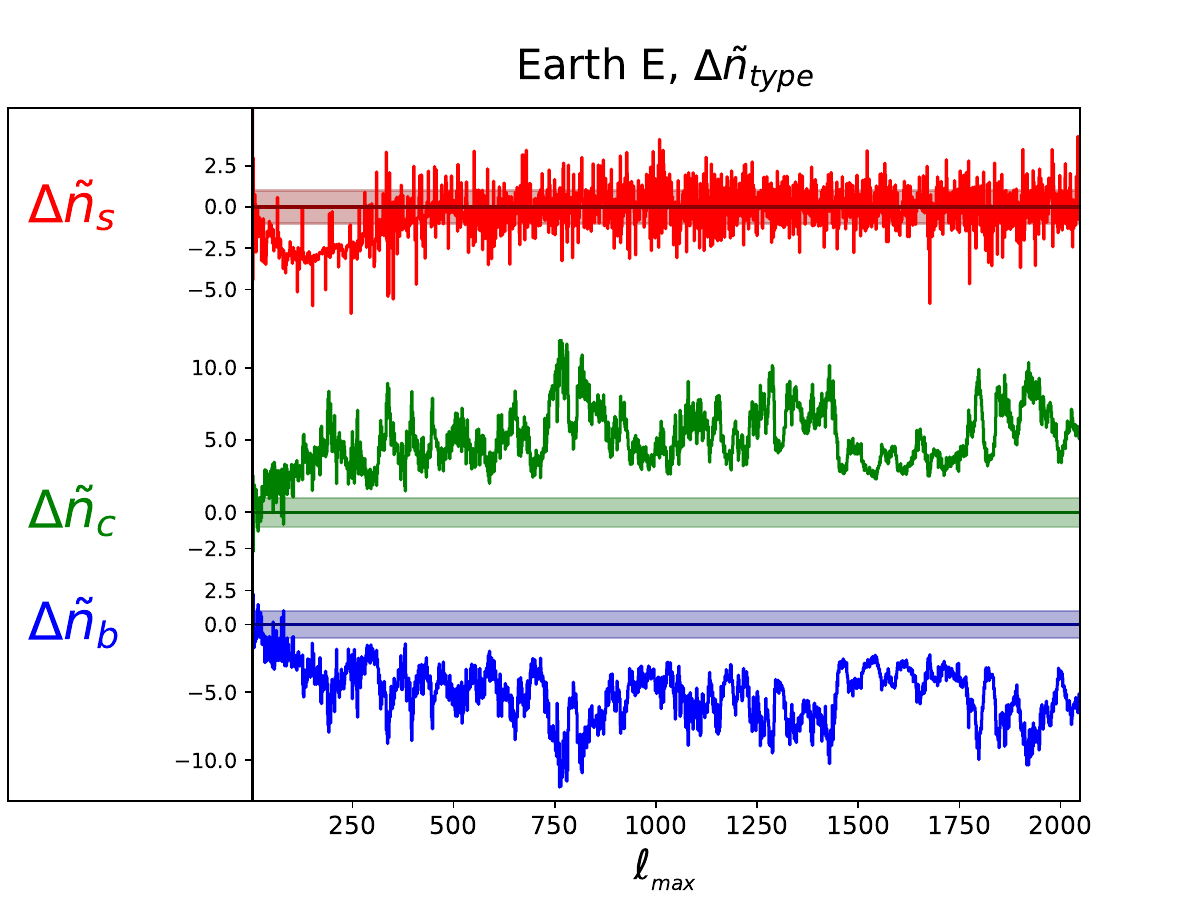}
  \includegraphics[width=0.49\textwidth]{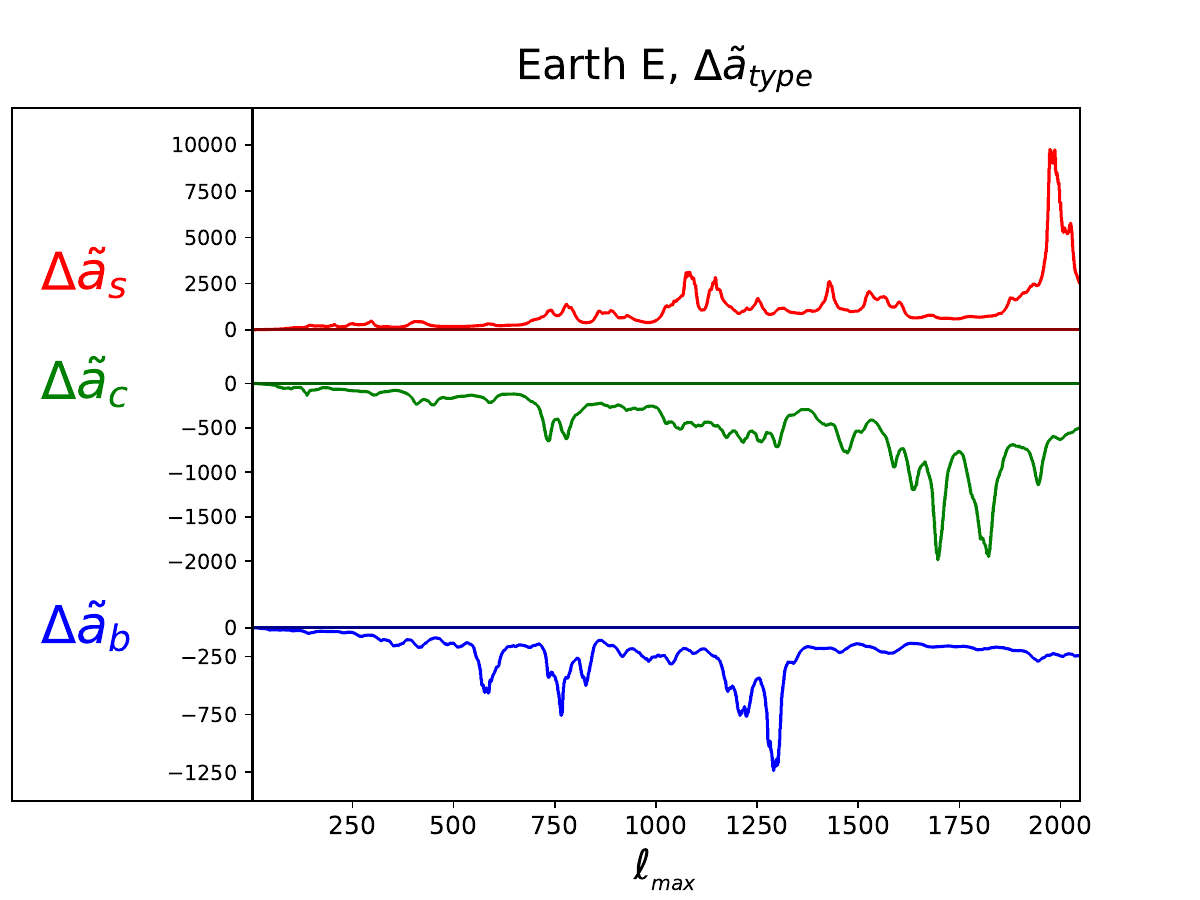}   
  \includegraphics[width=0.24\textwidth]{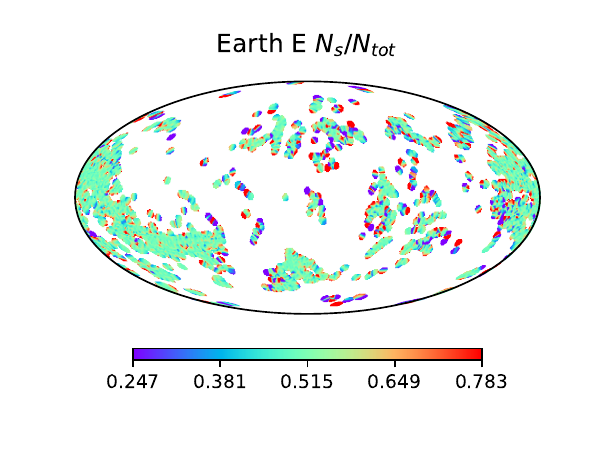}
   \includegraphics[width=0.24\textwidth]{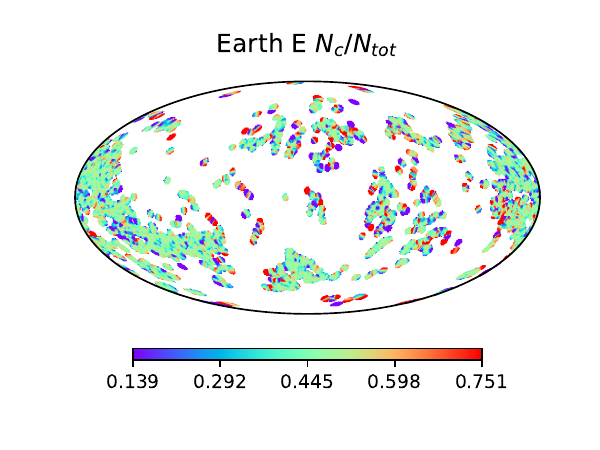}
   \includegraphics[width=0.24\textwidth]{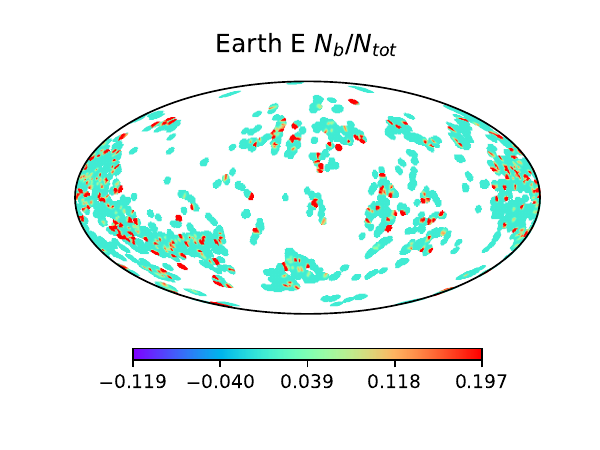}
   \includegraphics[width=0.24\textwidth]{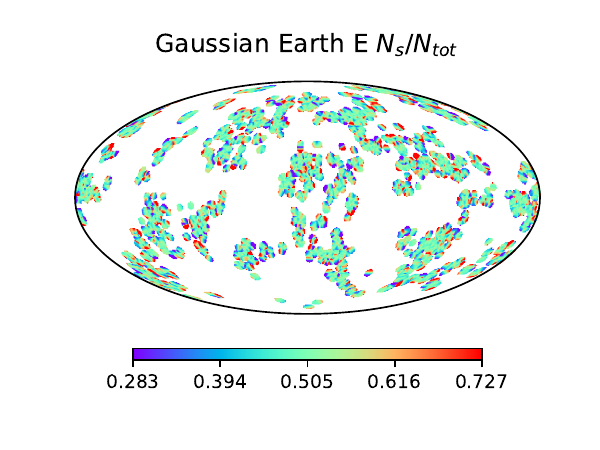}
   \includegraphics[width=0.24\textwidth]{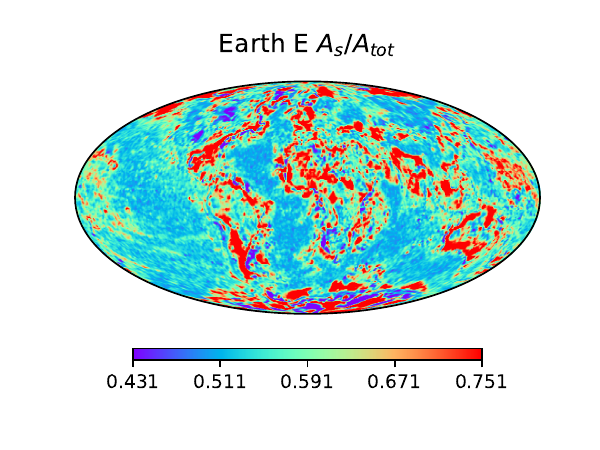}
   \includegraphics[width=0.24\textwidth]{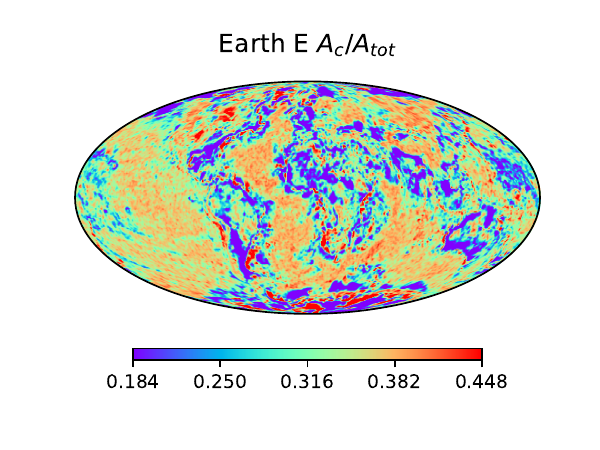}
   \includegraphics[width=0.24\textwidth]{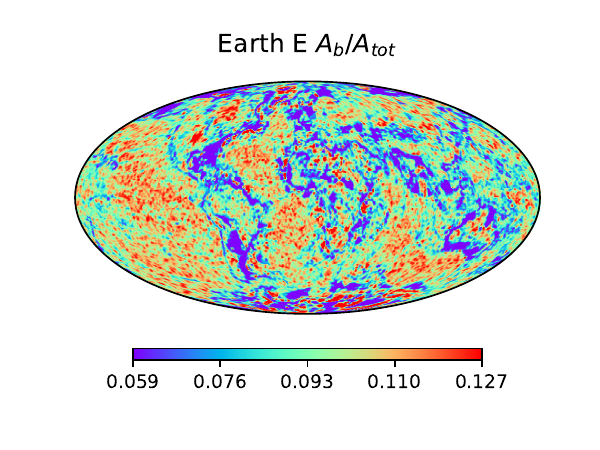}
   \includegraphics[width=0.24\textwidth]{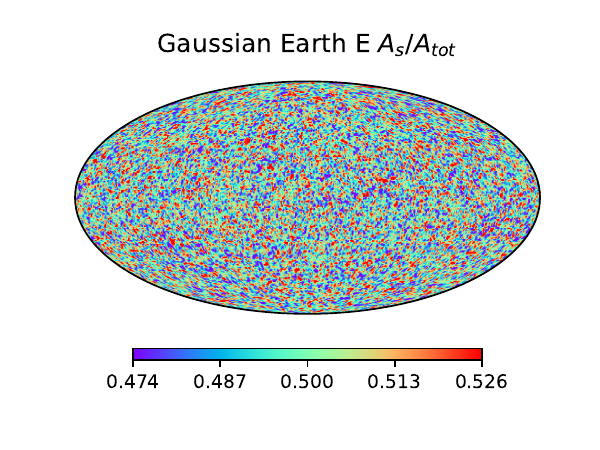}
   \includegraphics[width=0.24\textwidth]{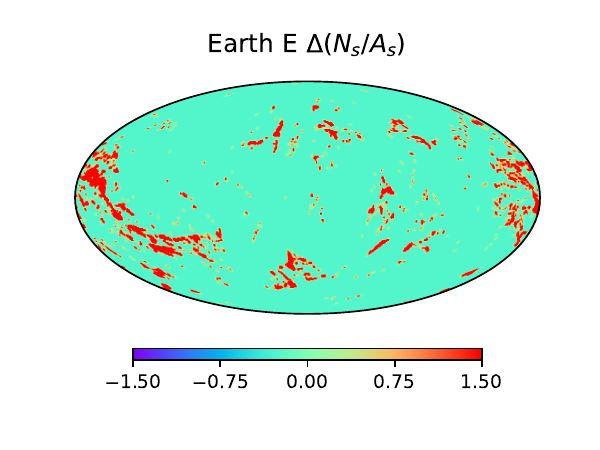}
   \includegraphics[width=0.24\textwidth]{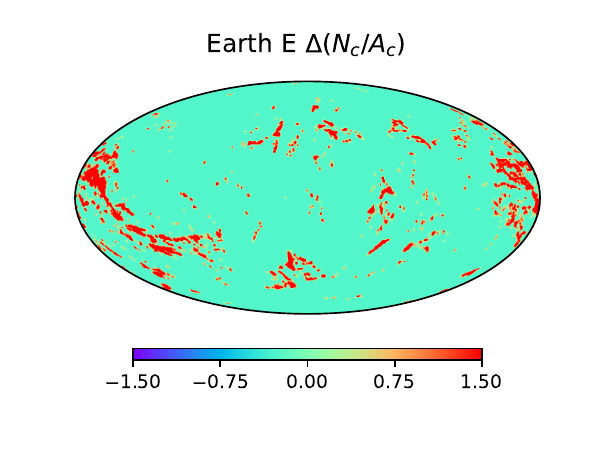}
   \includegraphics[width=0.24\textwidth]{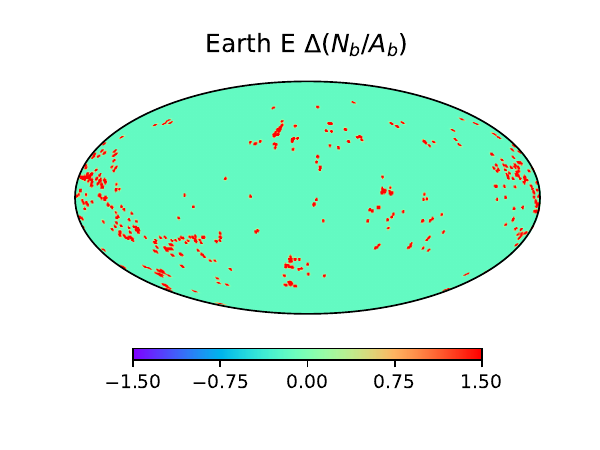}
   \includegraphics[width=0.24\textwidth]{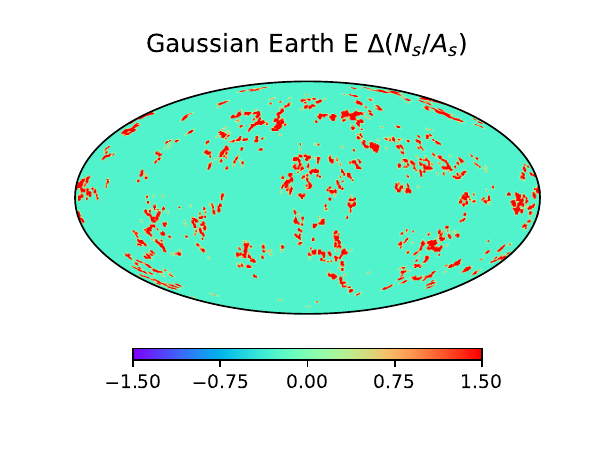}
   \caption{Statistics of singularities and area fractions for the E polarization of the Earth.
     {\it Upper left panel:} deviations of the fractions of points
     of three different types from the average expected values in a Gaussian field as a function of $\ell_{max}$:
     $\Delta\tilde{n}_{type}(\ell_{max})$. The shaded areas show $1\cdot\sigma$ deviations which correspond to
     $\langle(\Delta\tilde{n}_{type}^G)^2\rangle^{\frac{1}{2}}$.
     {\it Upper right panel:} the same as the left panel, but for the area fractions $\Delta\tilde{a}_{type}(\ell_{max})$
     (the shaded areas are very small).
     {\it The bottom three rows from top to bottom in descending order:}
     Maps showing the  relative concentrations of each singularity point type
     $\frac{N_{type}}{N_{tot}}$, their area fractions
     $\frac{A_{type}}{A_{tot}}$ and their spatial density
     $\Delta\left(\frac{N_{type}}{A_{type}}\right)$. For comparison, the right column shows such maps for
     saddles in the case of a Gaussian field with the Earth spectrum. The smoothing angle is
     $\theta_0=0.5^\circ$, which corresponds to $\sim$ 35 miles on the Earth's surface.
     The white part of the surface does not contain singularities.} 
\end{figure*}

Deviations from the expected average values (Fig. 3) significantly exceeding the standard deviation mean the
presence of a non-Gaussian component in the signal due to foregrounds,
gravitational lensing or initial non-Gaussianity. In particular, if such a deviation appears at some
$\ell_{max}$, then this may indicate the presence of an unremoved foreground in the map with a
characteristic angular scale $\theta\sim\pi/\ell_{max}$ .

\section{Planck data analysis}

In this Section we perform a detailed analysis of Planck CMB polarization maps (SMICA), first demonstrating a
strongly non-Gaussian example of E polarization caused by the scalar field of the Earth's surface.
In our studies, we used the method described in \citep{2025PhRvD.111f3536N} to find unpolarized
points and determine their types.

For fast simulation of $L$ maps with increasing number of harmonics $\ell_{max}=2,...,L$,
it is convenient to use a different spherical harmonic decomposition algorithm than HEALPix. Our approach
is based on the recurrent calculation of harmonics with a changing $m$ and a fixed $\ell$.
At each step we add harmonics  $a_{\ell m}Y_{\ell m}(\theta,\varphi)$ with $\ell=\ell_{max}+1$ and $m=-\ell,..,\ell$
to the already created signal consisting of harmonics with $2\le\ell\le\ell_{max}$ and $-\ell\le m\le\ell$.
As a result $O(L^3\log_{{}_2}L)$ operations are required to simulate $L$ maps. Using the HEALPix algorithm
directly would require $O(L^4)$ operations, since to create a single map with $2\le\ell\le\ell_{max}$ harmonics
HEALPix needs $O(\ell_{max}^3)$ operations.  A detailed description of our algorithm is given in the Appendix.

For convenience, below we use the following notation:
\begin{equation}
   \begin{array}{l}
     \vspace{0.2cm}
     \tilde{n}_{type}=\frac{N_{type}}{N_{tot}},\hspace{0.1cm}\tilde{a}_{type}=\frac{A_{type}}{A_{tot}},
     \hspace{0.1cm}\tilde{\rho}_{type}=\frac{N_{type}}{A_{type}},\\
     \vspace{0.2cm}
     \Delta\tilde{n}_{type}=\tilde{n}_{type}-\langle\tilde{n}_{type}^G\rangle,\\
     \vspace{0.2cm}
     \Delta\tilde{a}_{type}=\tilde{a}_{type}-\langle\tilde{a}_{type}^G\rangle,\\
     \Delta\tilde{\rho}_{type}=\tilde{\rho}_{type}-\langle\tilde{\rho}_{type}^G\rangle,
     \hspace{0.1cm}type =s,c,b.
   \end{array}
\end{equation}
Here $\tilde{x}_{type}=\tilde{x}_{type}(\ell_{max})$ denotes the observed value being analyzed, and
$\langle\tilde{x}_{type}^G\rangle$ denotes the
mean of the same value for a Gaussian field. If the observed field corresponds to Gaussian statistics,
then $\langle\Delta\tilde{x}_{type}\rangle=0$
and $\langle(\Delta\tilde{x}_{type})^2\rangle=\langle(\Delta\tilde{x}_{type}^G)^2\rangle$ is the square of the standard deviation of this
value for a given spectrum and the size of the investigated sky region $A_{tot}$.

\subsection{The Earth's surface as an example of non-Gaussianity}

The surface of our planet can be taken as an example of a scalar field
$E^{{}^{\boldsymbol{\oplus}}}(\theta,\varphi)$ on a sphere, which by its nature is very
far from Gaussian. In terms of spherical harmonics the map of the Earth is,
\begin{equation}
   \begin{array}{l}
     \vspace{0.2cm}
     E^{{}^{\boldsymbol{\oplus}}}=\sum\limits_{\ell=0}^{\ell_{max}}\sum\limits_{m=-\ell}^\ell a_{\ell m}^{{}^{\boldsymbol{\oplus}}}
     \cdot Y_{\ell m}(\boldsymbol{\eta}),
   \end{array}
\end{equation}
where $\ell_{max}$ denotes the maximum number of harmonics used, which is equivalent to the resolution
or the minimum size of the examined relief features.
Let us now consider the E polarization field on a sphere whose source is the
scalar $E^{{}^{\boldsymbol{\oplus}}}$. In this case, the Stokes parameters at any point on
the map are obtained by taking the covariant second derivatives of
$E^{{}^{\boldsymbol{\oplus}}}$ with respect to x and y [see Eqs. (5),(6)] or, equivalently, they can be
written in terms of spin-weighted spherical harmonics \citep{1967JMP.....8.2155G, 1997PhRvL..78.2054S},
\begin{equation}
  \begin{array}{l}
    \vspace{0.3cm}
    Q^{{}^{\boldsymbol{\oplus}}}
    =\frac{1}{2}\sum\limits_{\ell,m}
   \sqrt{\frac{(\ell+2)!}{(\ell-2)!}}a_{\ell m}^{{}^{\boldsymbol{\oplus}}}
    \left[{}_{2}Y_{\ell m}(\boldsymbol{\eta})+
      {}_{-2}Y_{\ell m}(\boldsymbol{\eta})\right], \\
    U^{{}^{\boldsymbol{\oplus}}}
    =\frac{1}{2i}\sum\limits_{\ell,m}
    \sqrt{\frac{(\ell+2)!}{(\ell-2)!}}a_{\ell m}^{{}^{\boldsymbol{\oplus}}}
    \left[{}_{2}Y_{\ell m}(\boldsymbol{\eta})-
         {}_{-2}Y_{\ell m}(\boldsymbol{\eta})\right].
    \end{array}
\end{equation}
The non-Gaussian statistics of the polarization described by Eq. (19) arise due to
a very specific correlation between the Earth coefficients $a_{\ell m}^{{}^{\boldsymbol{\oplus}}}$.
For a random Gaussian field $\widetilde{E}^{{}^{\boldsymbol{\oplus}}}$ with identical power
spectrum $\widetilde{C}_\ell^{{}^{\boldsymbol{\oplus}}}=C_\ell^{{}^{\boldsymbol{\oplus}}}$,
the coefficients $\widetilde{a}_{\ell m}^{{}^{\boldsymbol{\oplus}}}$ must be independent
random Gaussian values with the following properties:
\begin{equation}
  \begin{array}{l}
    \vspace{0.2cm}
    \left<\widetilde{a}_{\ell m}^{{}^{\boldsymbol{\oplus}}}\right>=0,\hspace{0.2cm}
    \left<\widetilde{a}_{\ell m}^{{}^{\boldsymbol{\oplus}}}\cdot
    \widetilde{a}_{\ell m}^{{}^{\boldsymbol{\oplus}}\mathlarger{\mathlarger{\mathlarger{\ast}}}}\right>=
    C_\ell^{{}^{\boldsymbol{\oplus}}},\hspace{0.2cm}-\ell\le m\le\ell,\\
    \sum\limits_{m=-\ell}^{\ell}
    \widetilde{a}_{\ell m}^{{}^{\boldsymbol{\oplus}}}\cdot
    \widetilde{a}_{\ell m}^{{}^{\boldsymbol{\oplus}}\mathlarger{\mathlarger{\mathlarger{\ast}}}}
    \equiv\sum\limits_{m=-\ell}^{\ell}
    a_{\ell m}^{{}^{\boldsymbol{\oplus}}}\cdot
    a_{\ell m}^{{}^{\boldsymbol{\oplus}}\mathlarger{\mathlarger{\mathlarger{\ast}}}},
    \end{array}
\end{equation}
where $\langle\rangle$ means the ensemble average. Substituting the coefficients
$\widetilde{a}_{\ell m}^{{}^{\boldsymbol{\oplus}}}$ into Eq. (19), we obtain a Gaussian polarization field
with the terrestrial power spectrum. Comparison of the statistical properties of the real
``Earth polarization" with the polarization caused by a scalar Gaussian random field is
completely correct only if the spectrum of this field is identical to the spectrum of the Earth.
In this way, we make the signal phases random, while preserving the signal spectral properties.

\begin{figure*}[!htbp]
  \includegraphics[width=0.32\textwidth]{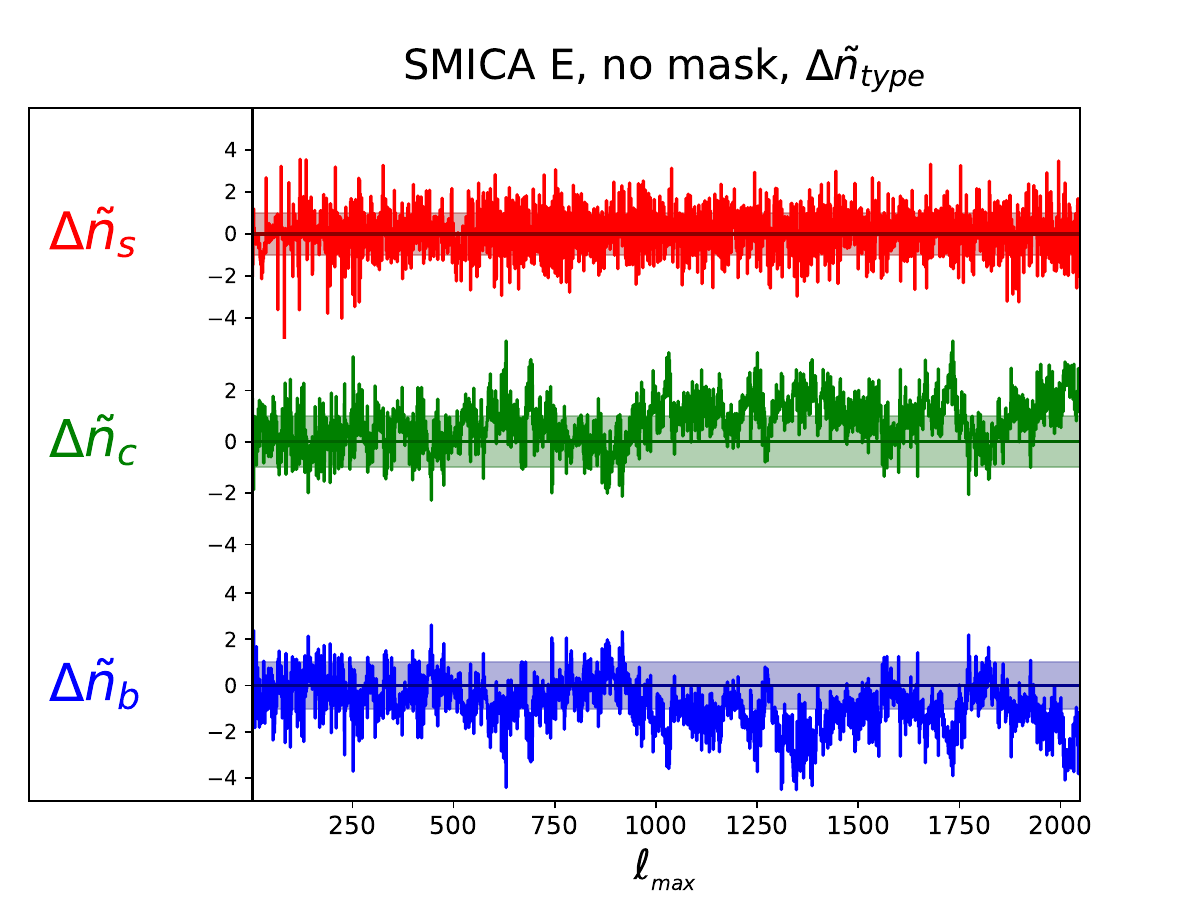}
  \includegraphics[width=0.32\textwidth]{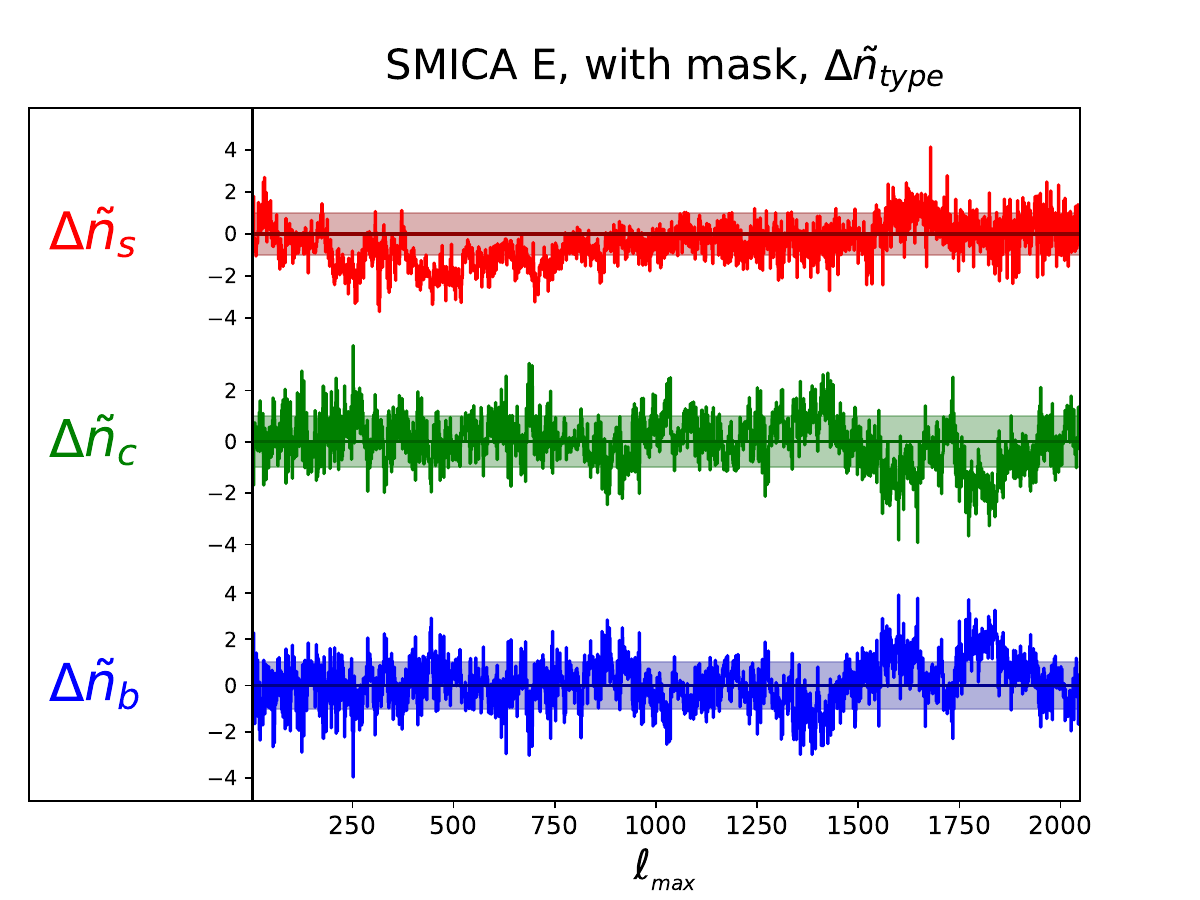}
  \includegraphics[width=0.32\textwidth]{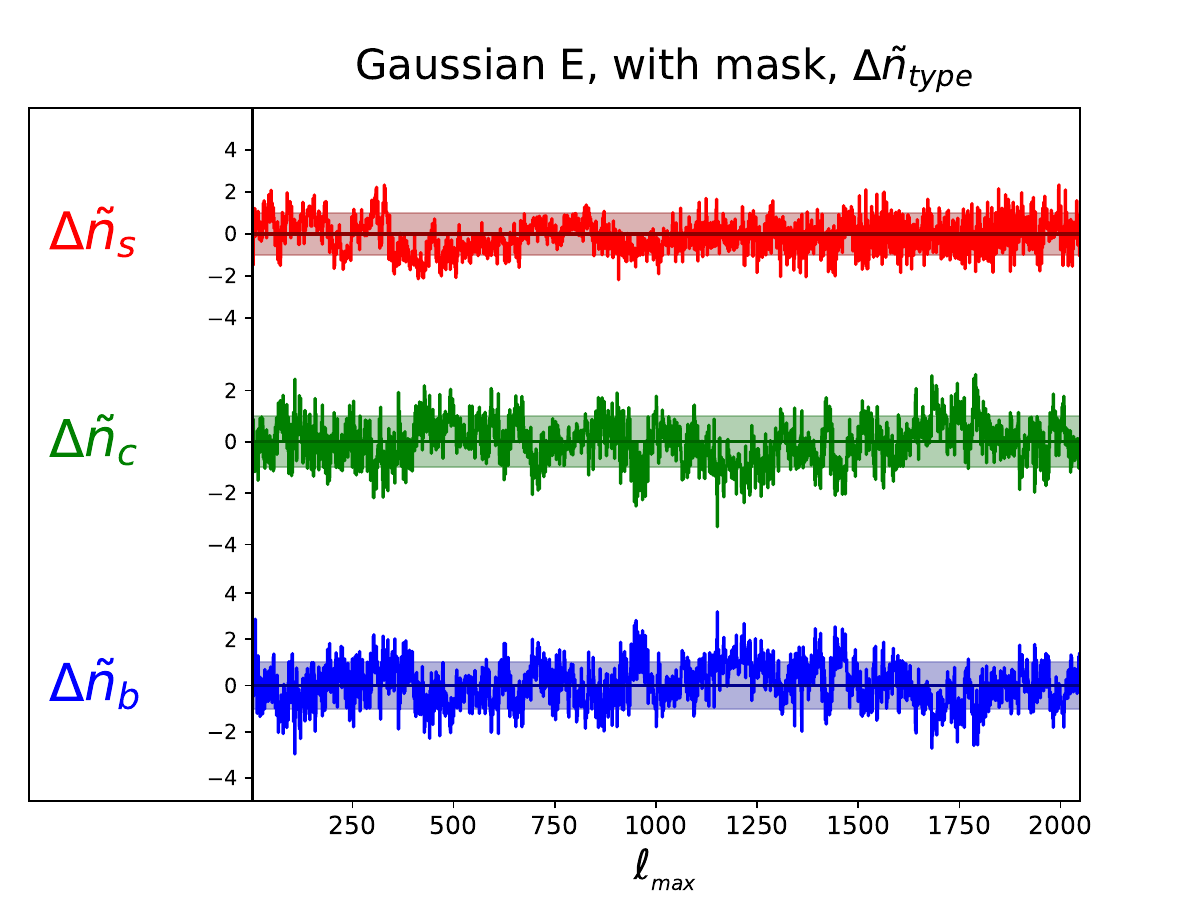} 
  \includegraphics[width=0.32\textwidth]{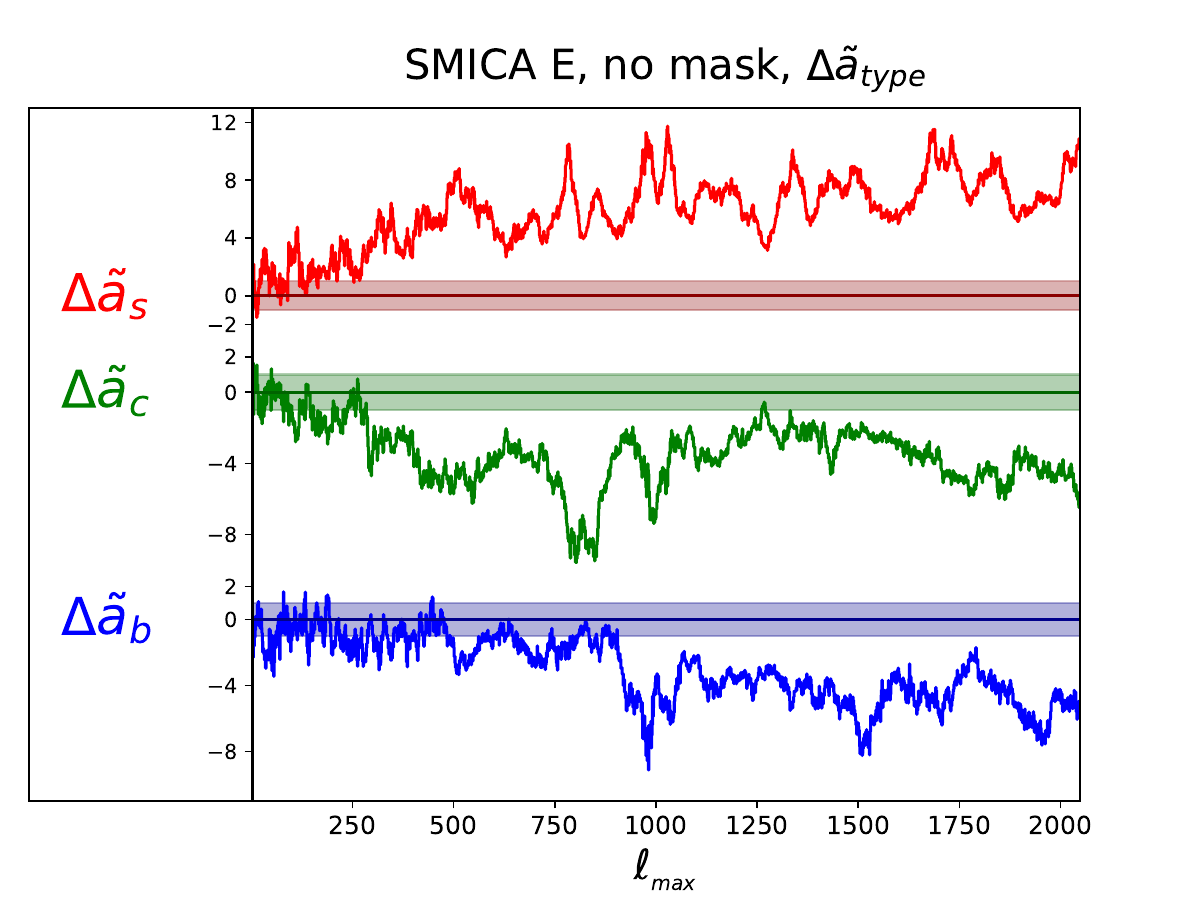}
  \includegraphics[width=0.32\textwidth]{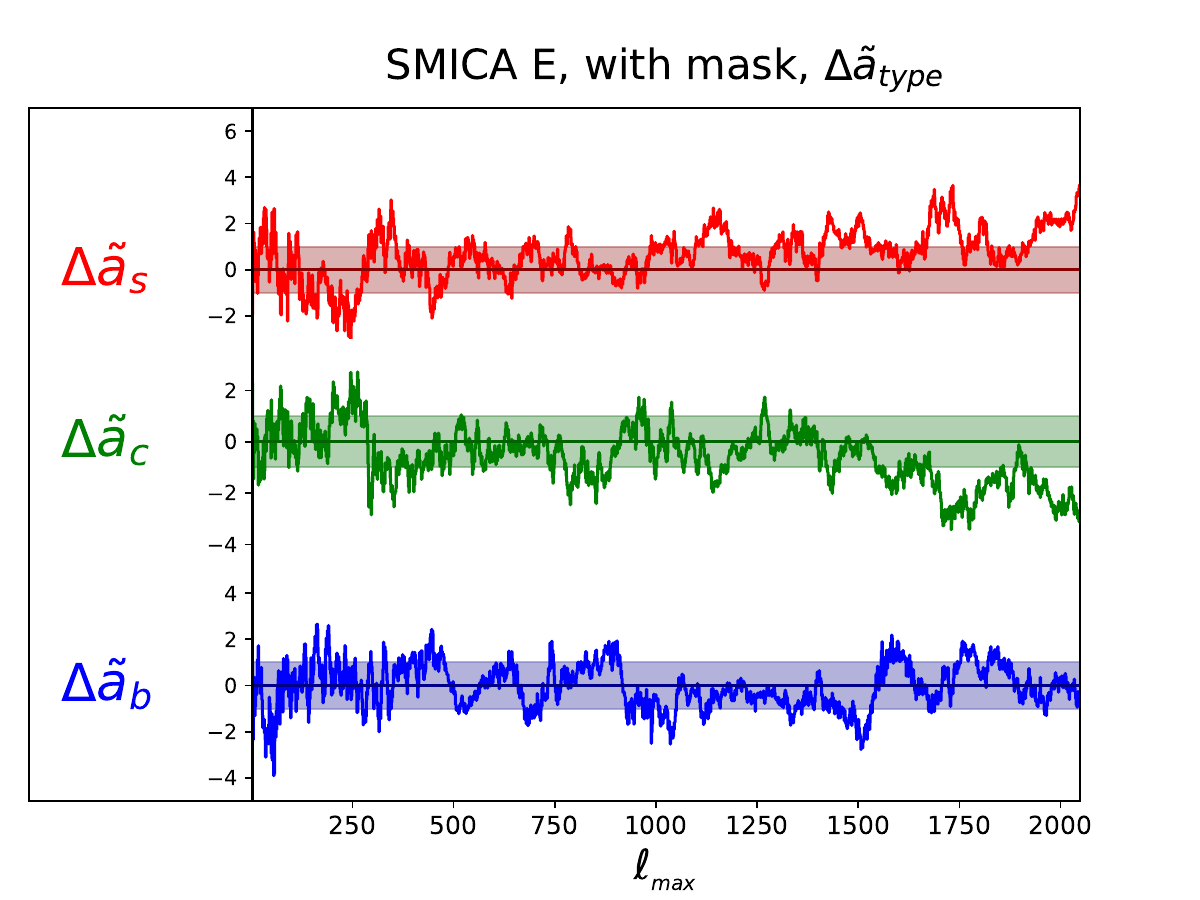}
  \includegraphics[width=0.32\textwidth]{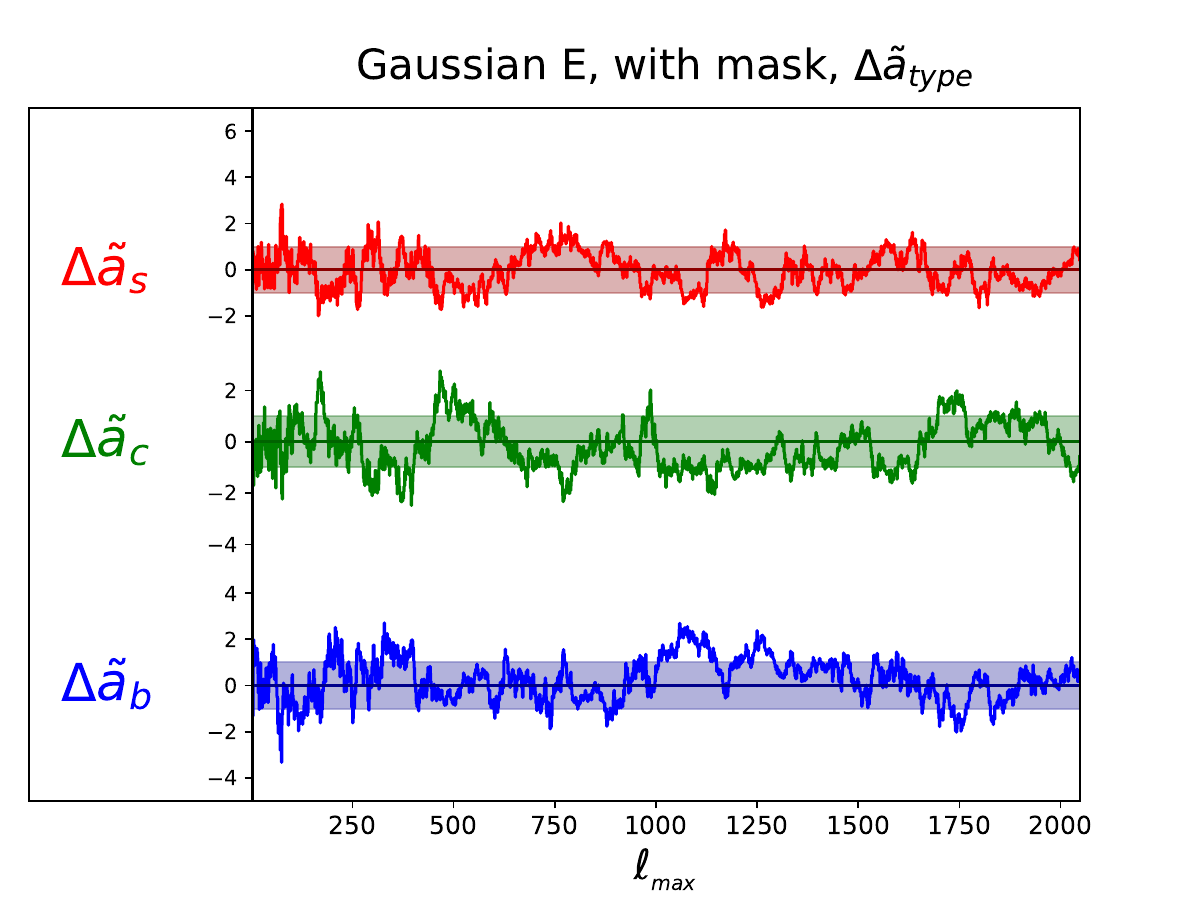}
   \includegraphics[width=0.32\textwidth]{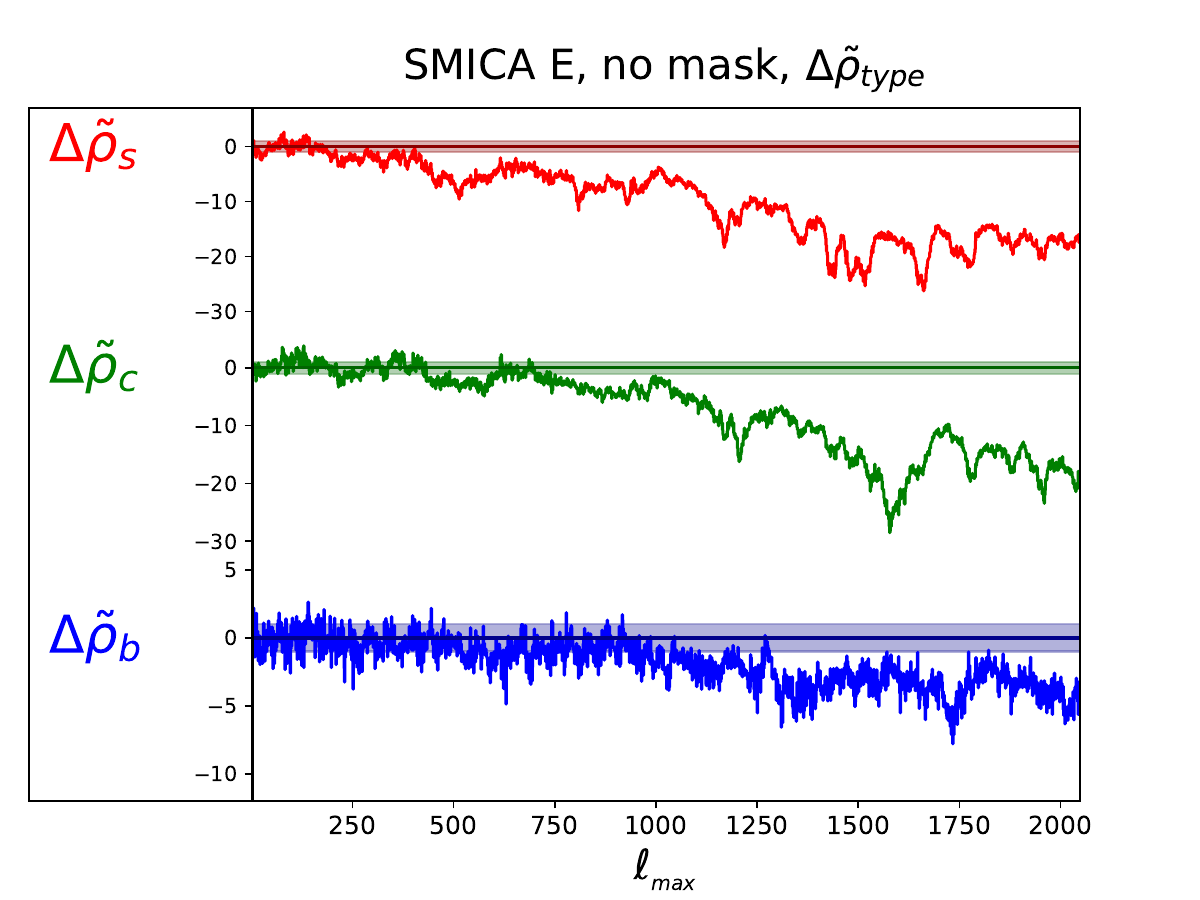}
  \includegraphics[width=0.32\textwidth]{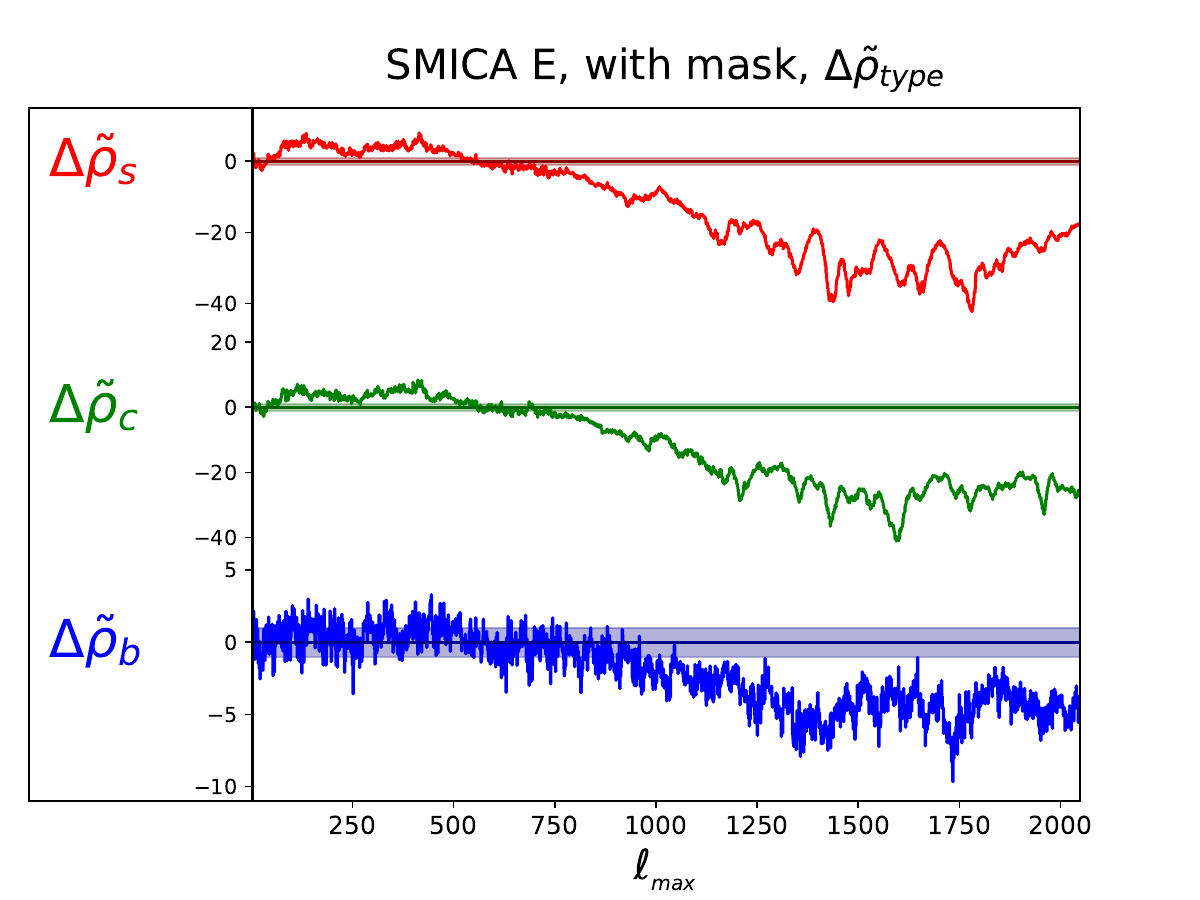}
  \includegraphics[width=0.32\textwidth]{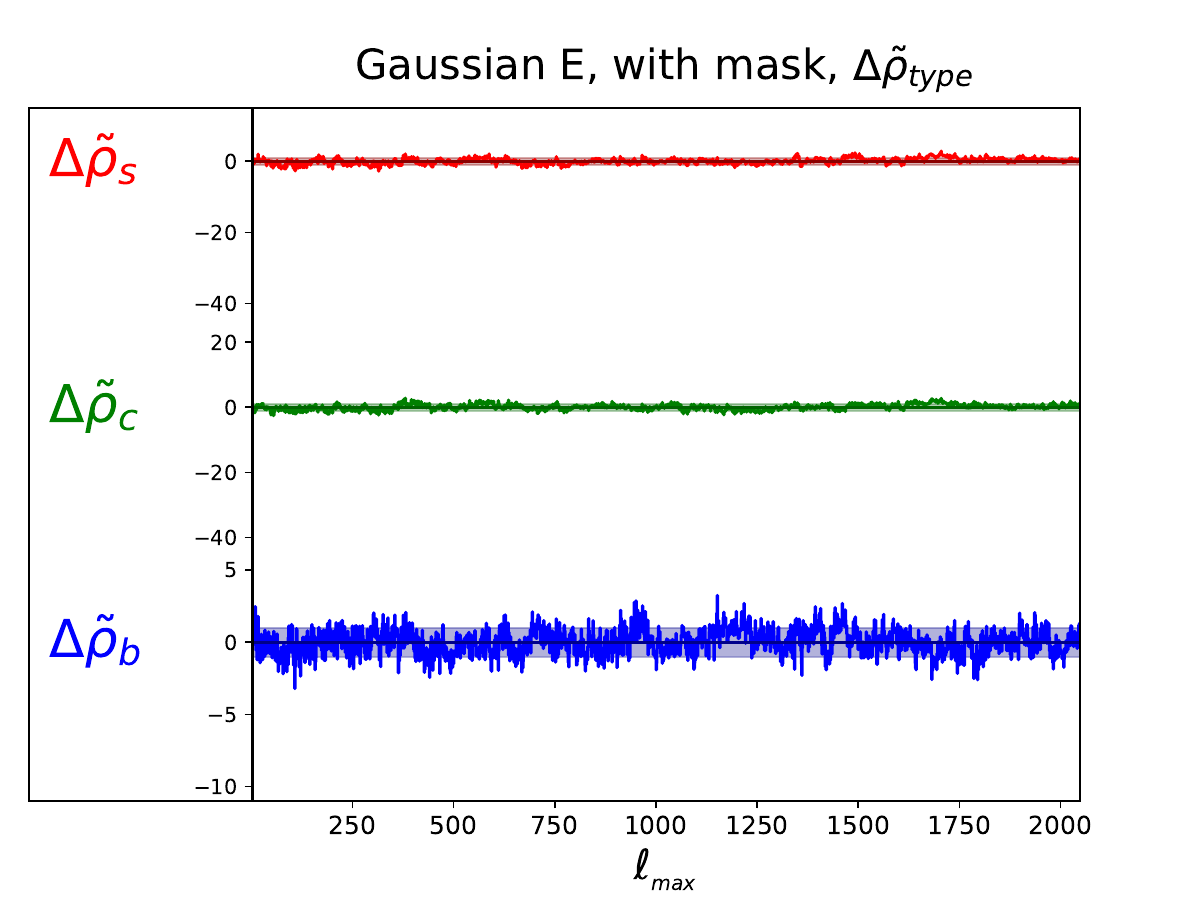}
  \includegraphics[width=0.24\textwidth]{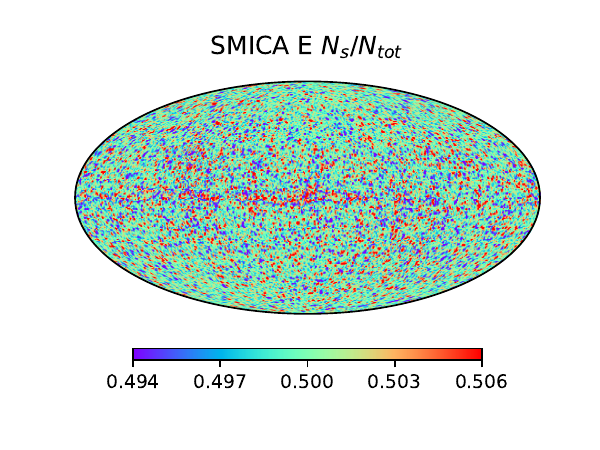}
   \includegraphics[width=0.24\textwidth]{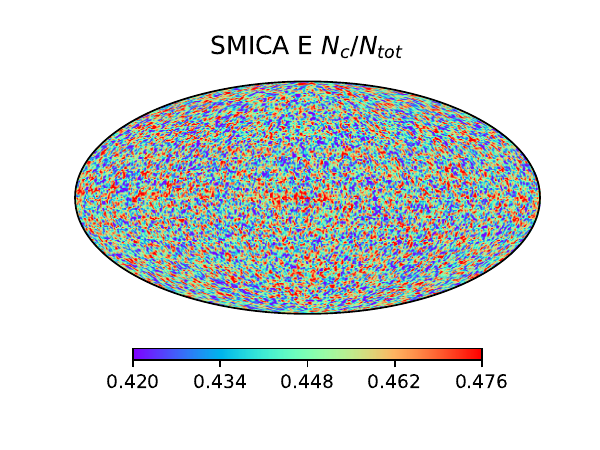}
   \includegraphics[width=0.24\textwidth]{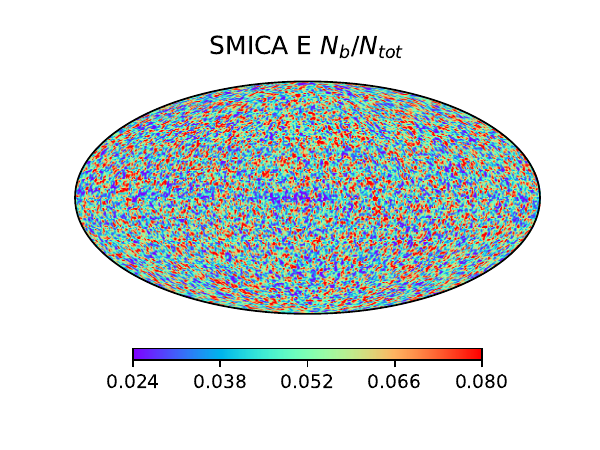}
   \includegraphics[width=0.24\textwidth]{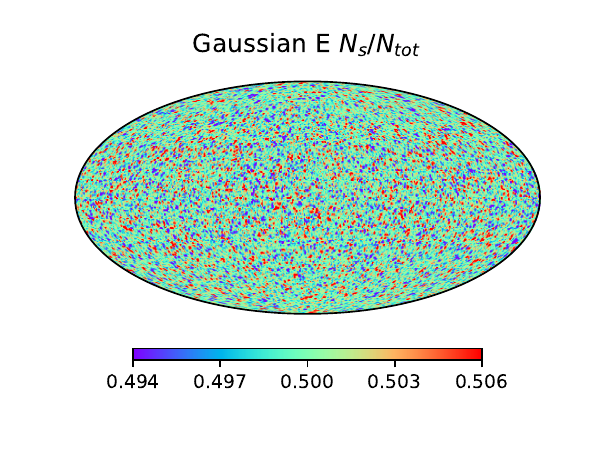}
   \includegraphics[width=0.24\textwidth]{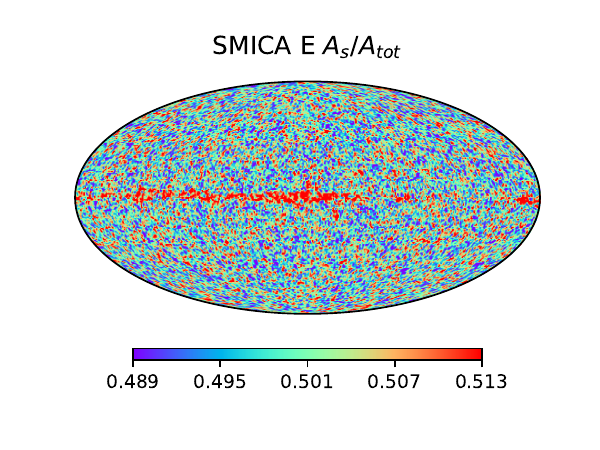}
   \includegraphics[width=0.24\textwidth]{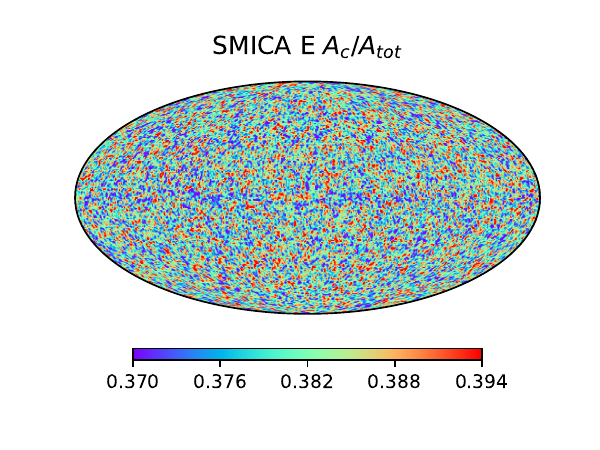}
   \includegraphics[width=0.24\textwidth]{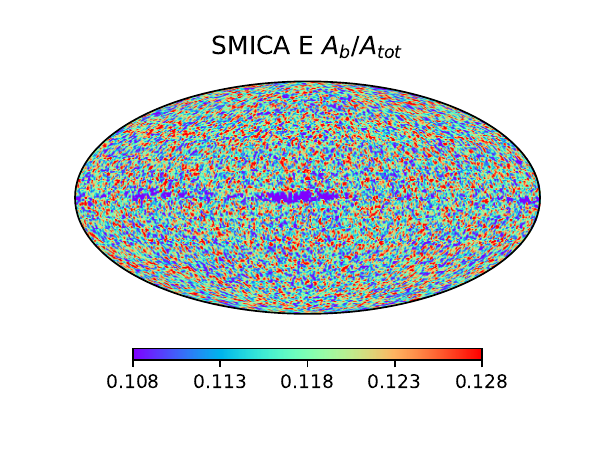}
   \includegraphics[width=0.24\textwidth]{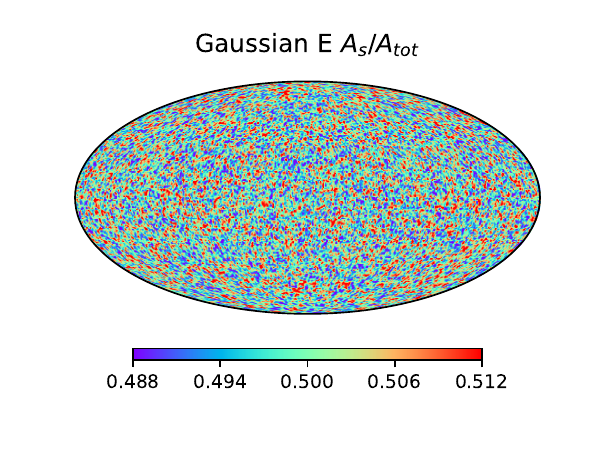}
   \includegraphics[width=0.24\textwidth]{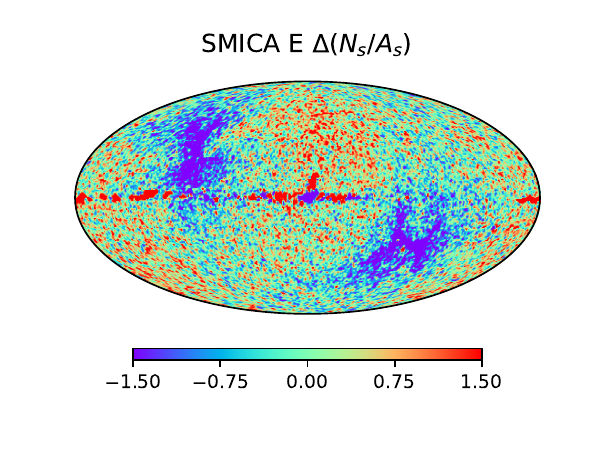}
   \includegraphics[width=0.24\textwidth]{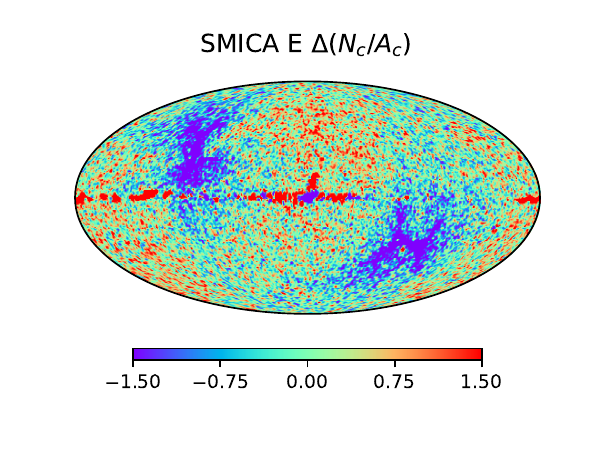}
   \includegraphics[width=0.24\textwidth]{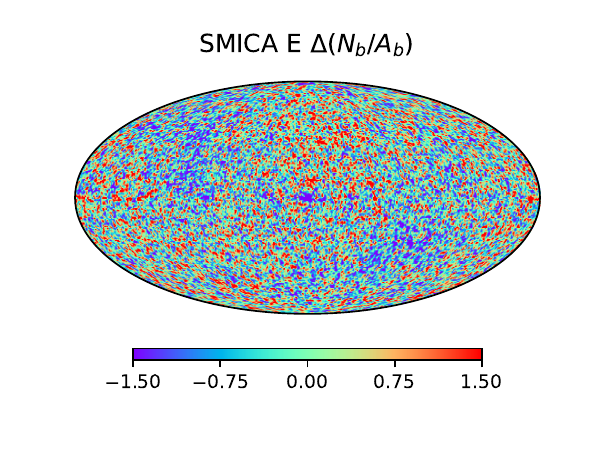}
   \includegraphics[width=0.24\textwidth]{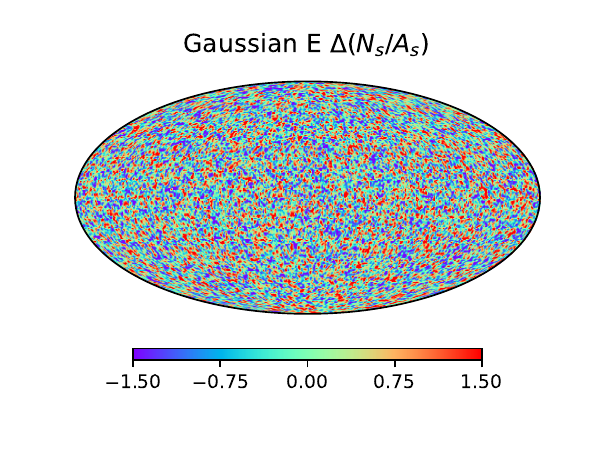}
   \caption{Statistics of the three singularity types (saddles, comets, beaks) and their corresponding area fractions in the SMICA CMB E mode polarization. {\it The first three rows (from top to bottom):} Deviations from the Gaussian mean
    for the relative number of singularities $\Delta\tilde{n}_{type}(\ell_{max})$, their area fractions
    $\Delta\tilde{a}_{type}(\ell_{max})$, and their spatial density $\Delta\tilde{\rho}_{type}(\ell_{max})$ in units of
    $\langle(\Delta\tilde{n}_{type}^G)^2\rangle^{\frac{1}{2}}$,
    $\langle(\Delta\tilde{a}_{type}^G)^2\rangle^{\frac{1}{2}}$ and  $\langle(\Delta\tilde{\rho}_{type}^G)^2\rangle^{\frac{1}{2}}$,
    respectively. The shaded areas mark the $1\sigma$ confidence region.
{\it  The last three rows (from top to bottom):}
Sky maps ($\ell_{max}=2048$, smoothed at $\theta_0 = 0.5^\circ$) showing the  relative concentrations of each singularity point type $\frac{N_{type}}{N_{tot}}$, the area fraction $\frac{A_{type}}{A_{tot}}$, and their spatial density
$\Delta\left(\frac{N_{type}}{A_{type}}\right)$. For comparison, the right column shows the same analysis performed on a typical Gaussian realization with the CMB SMICA E mode spectrum for saddles.}
\end{figure*}

\begin{figure*}[!htbp]
   \includegraphics[width=0.32\textwidth]{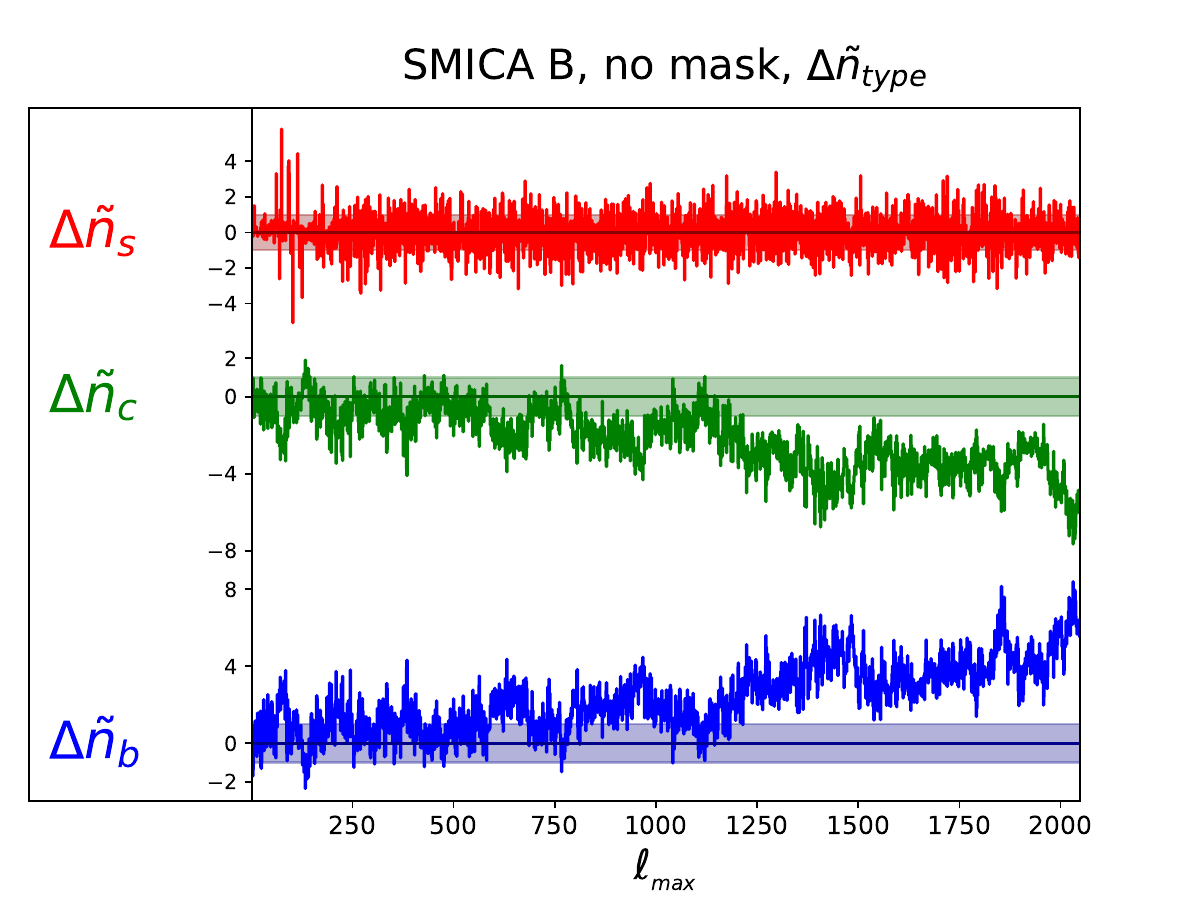}
   \includegraphics[width=0.32\textwidth]{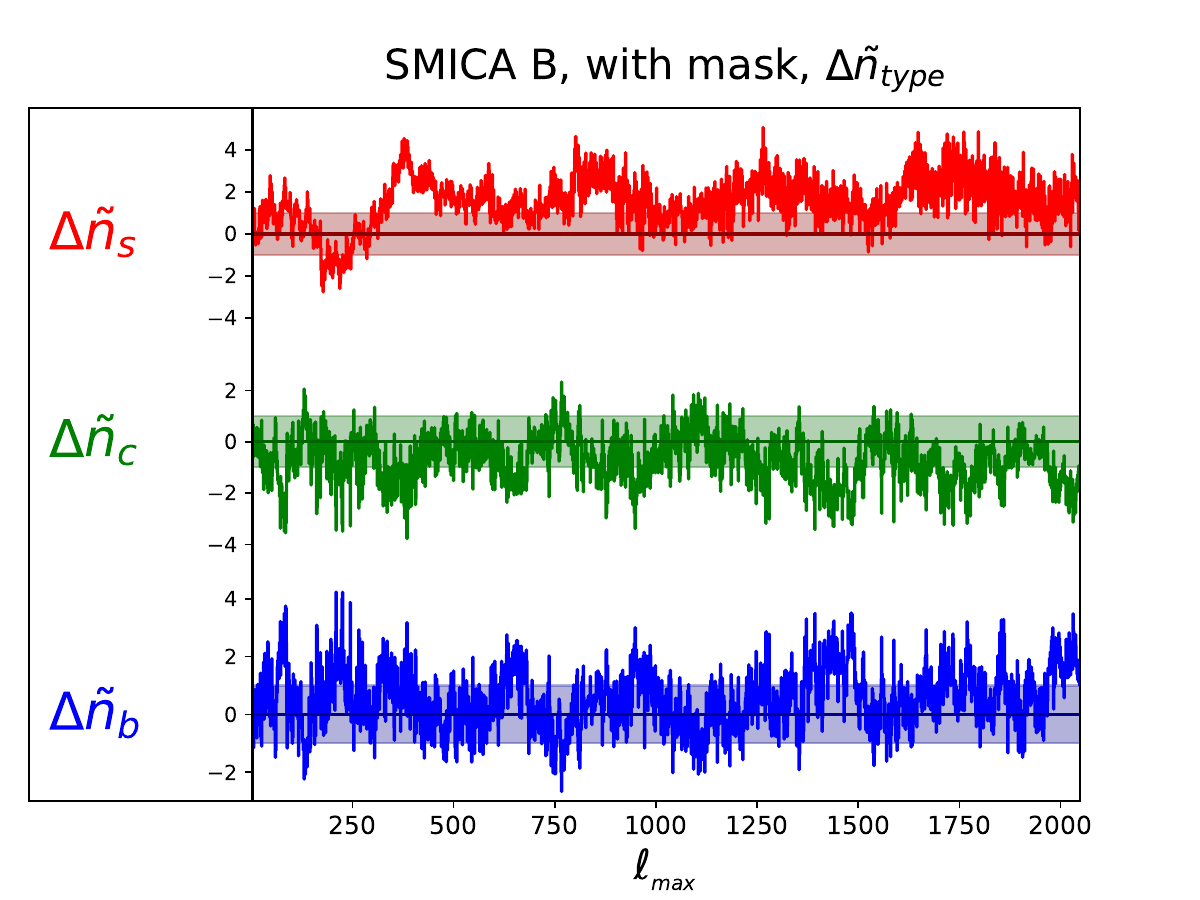}
   \includegraphics[width=0.32\textwidth]{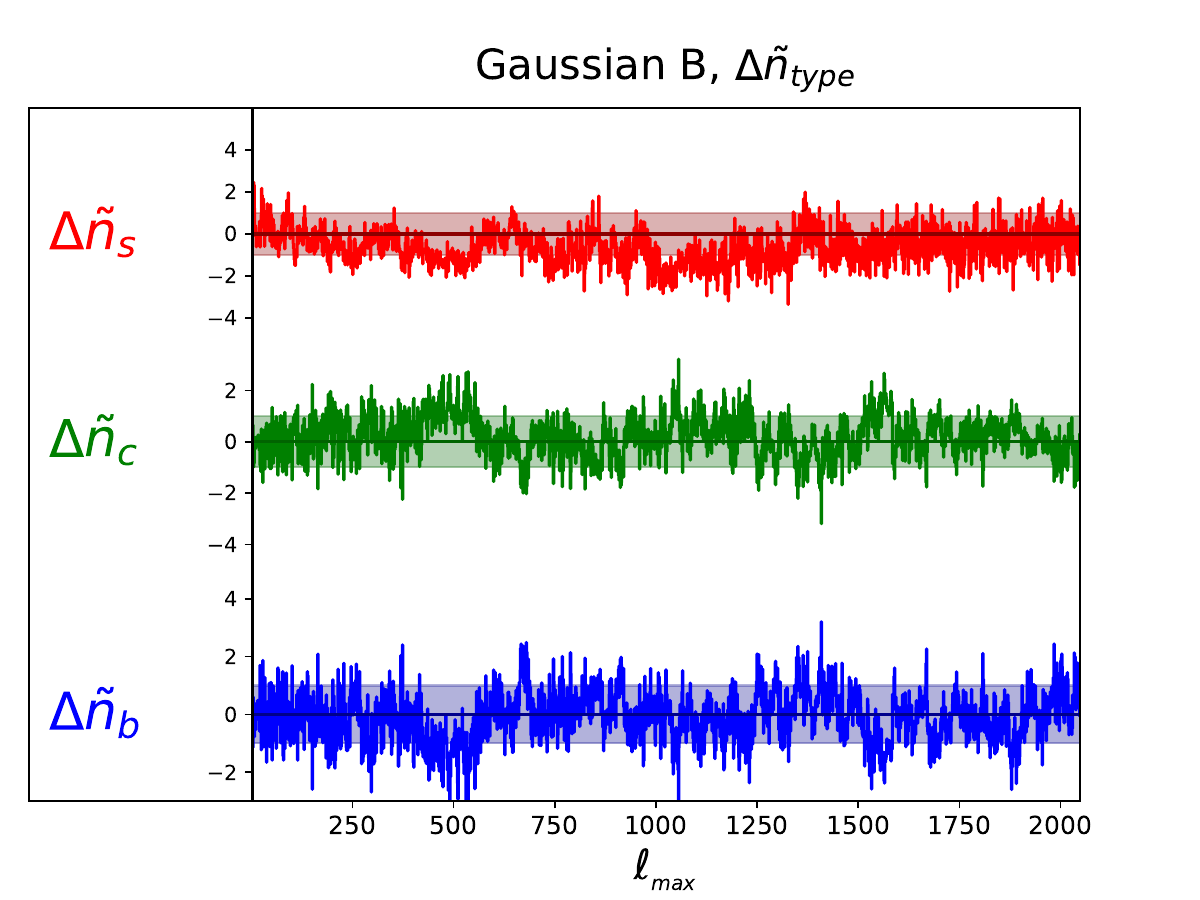}
   \includegraphics[width=0.32\textwidth]{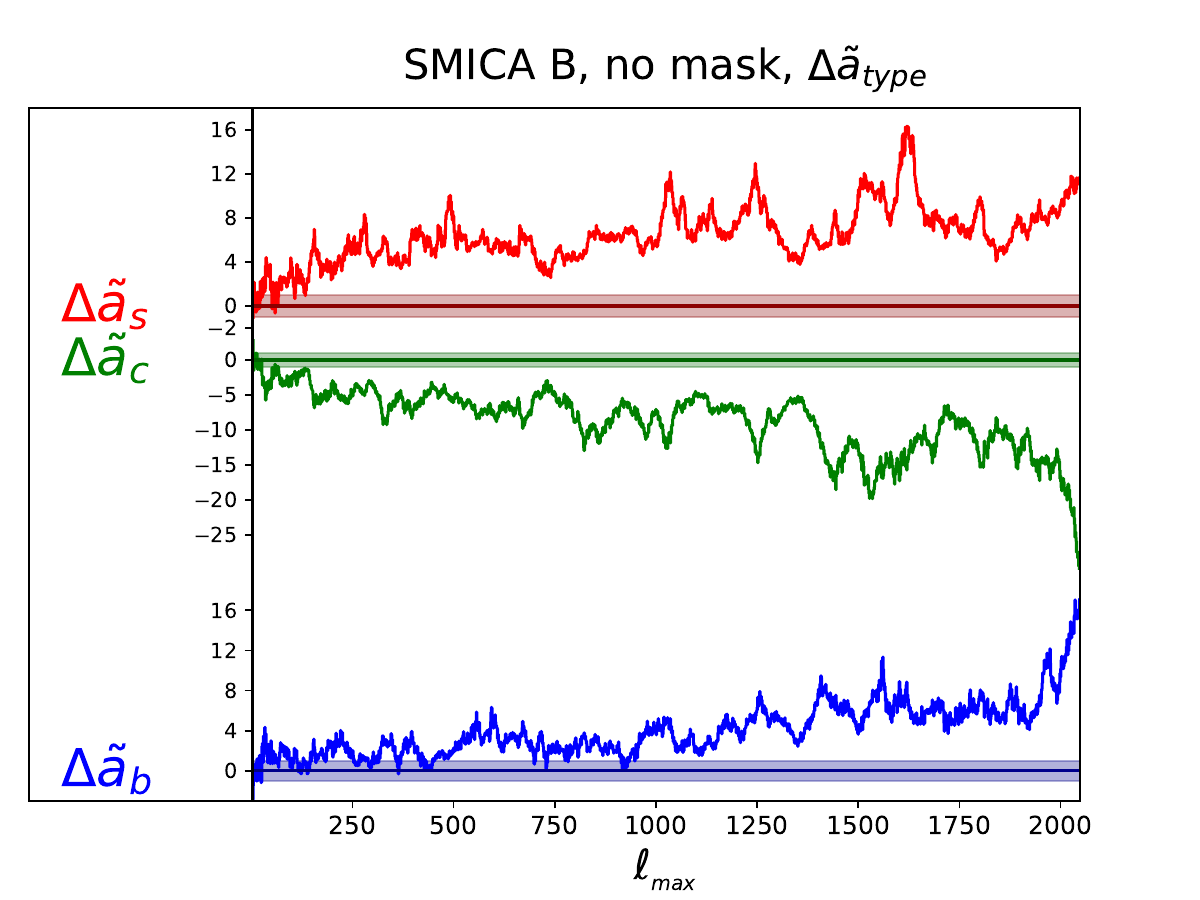}   
   \includegraphics[width=0.32\textwidth]{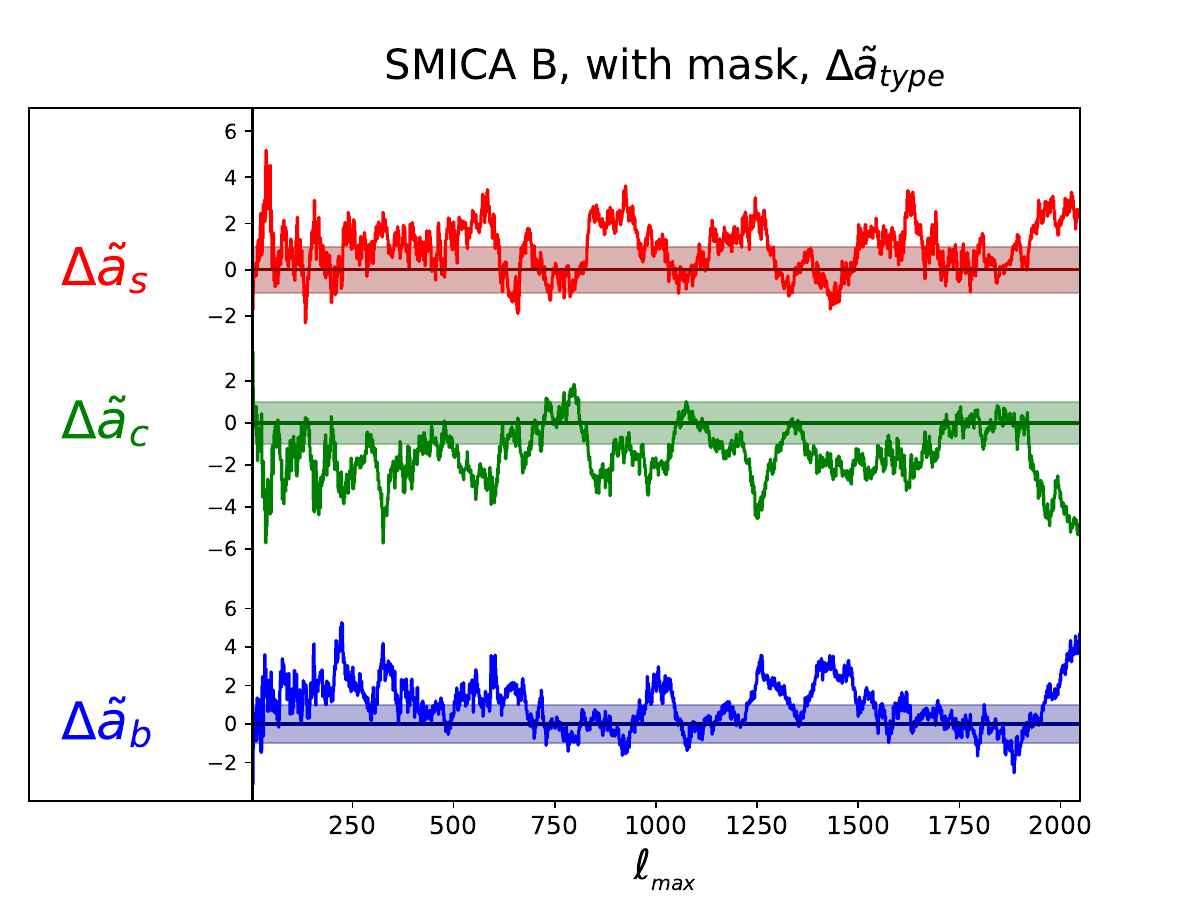}
   \includegraphics[width=0.32\textwidth]{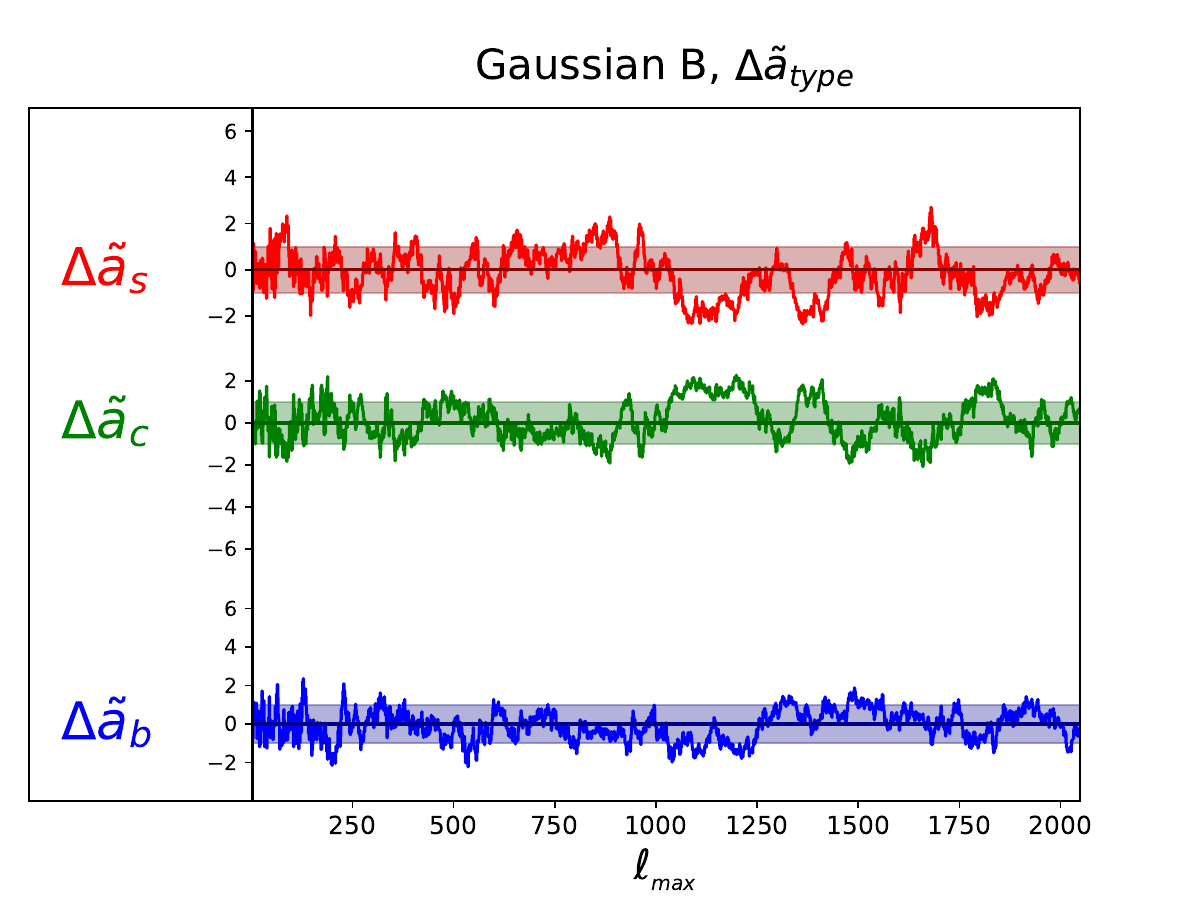}
   \includegraphics[width=0.32\textwidth]{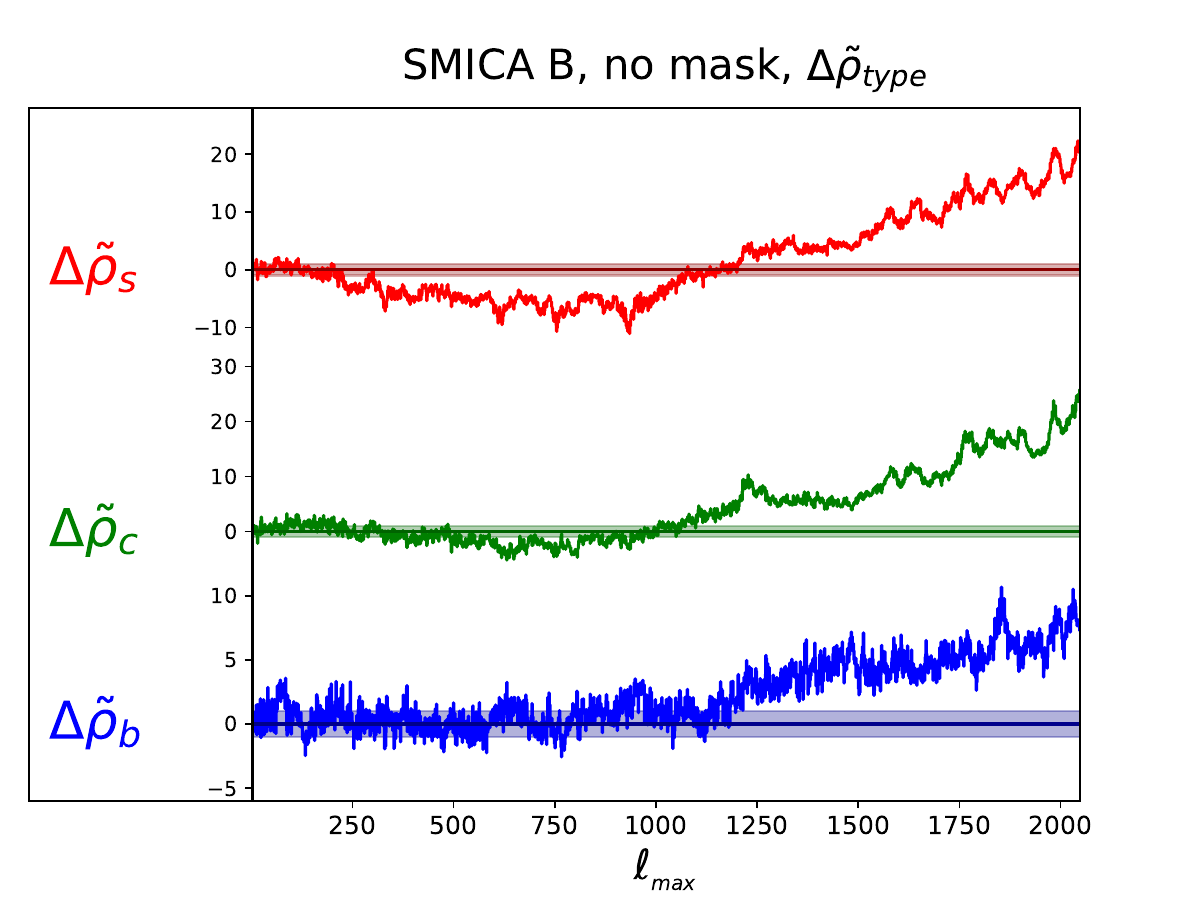}
   \includegraphics[width=0.32\textwidth]{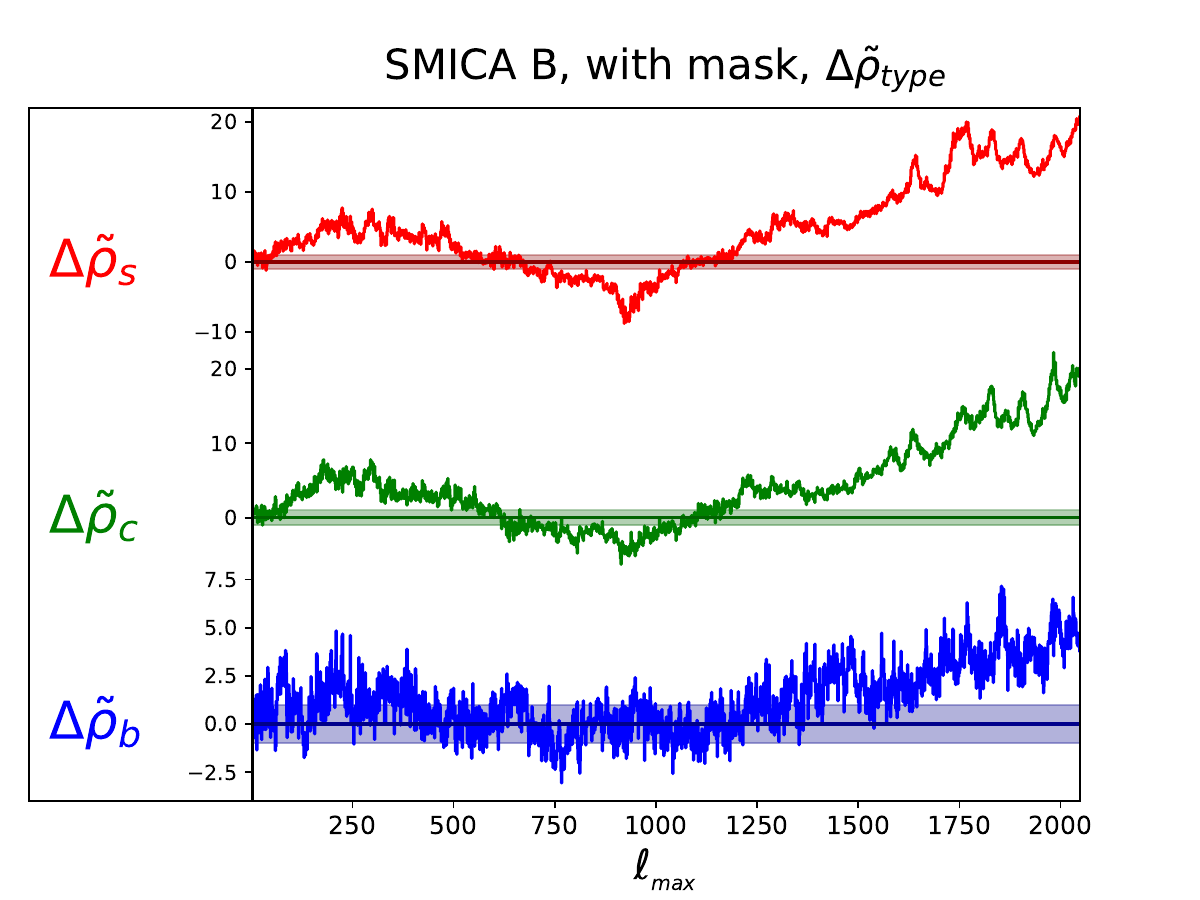}
   \includegraphics[width=0.32\textwidth]{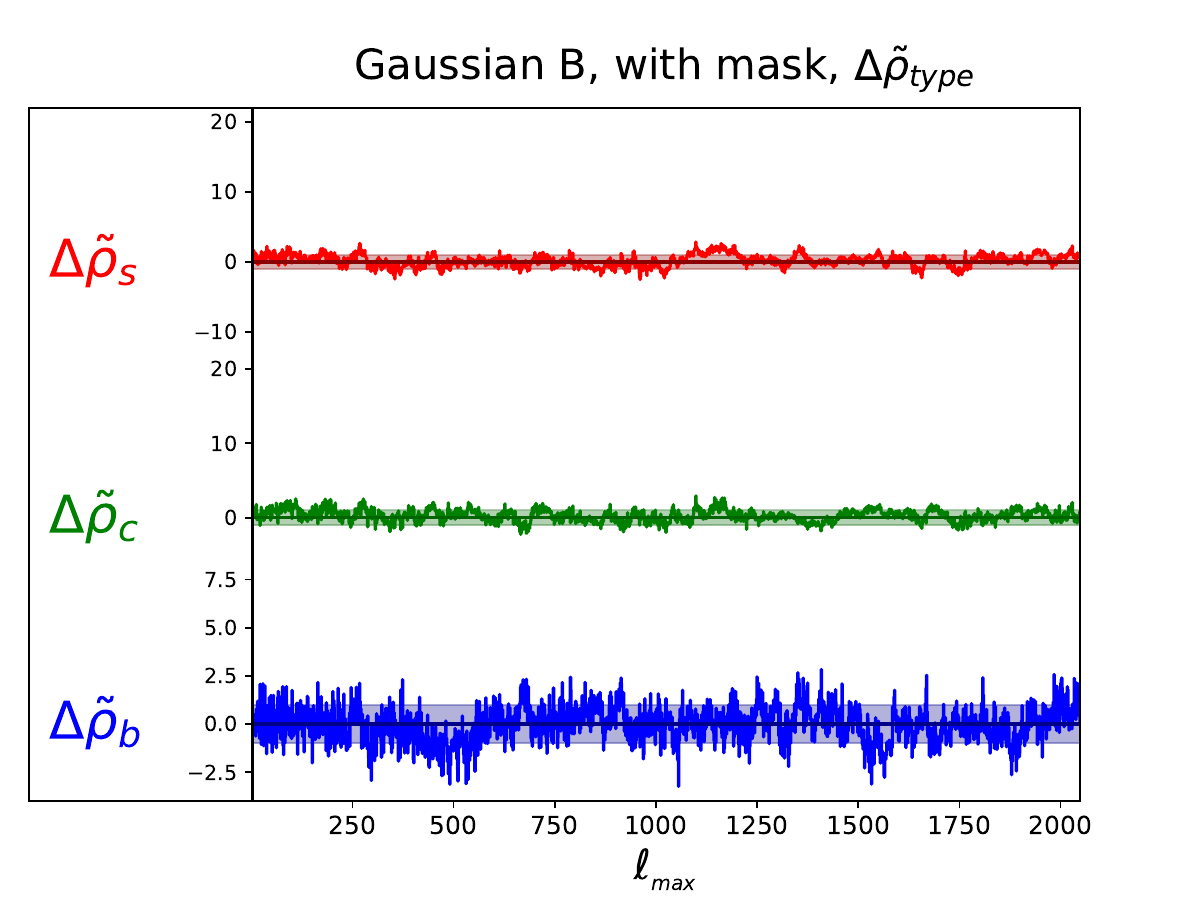}
   \includegraphics[width=0.24\textwidth]{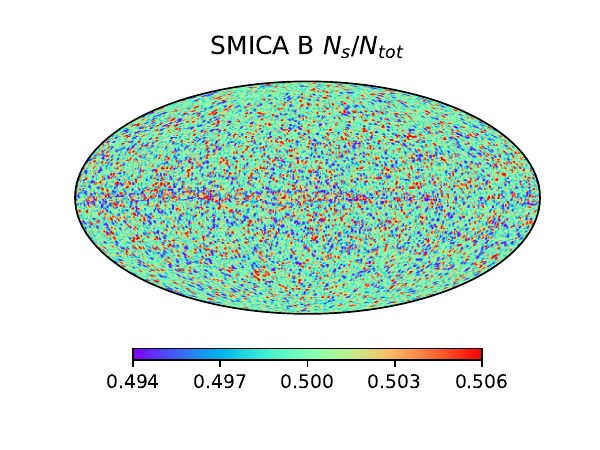}
   \includegraphics[width=0.24\textwidth]{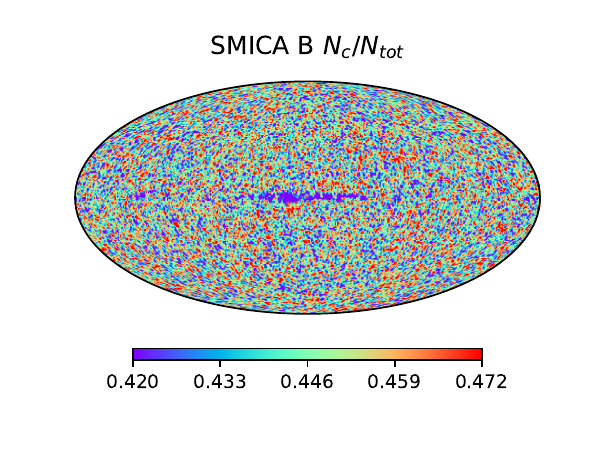}
   \includegraphics[width=0.24\textwidth]{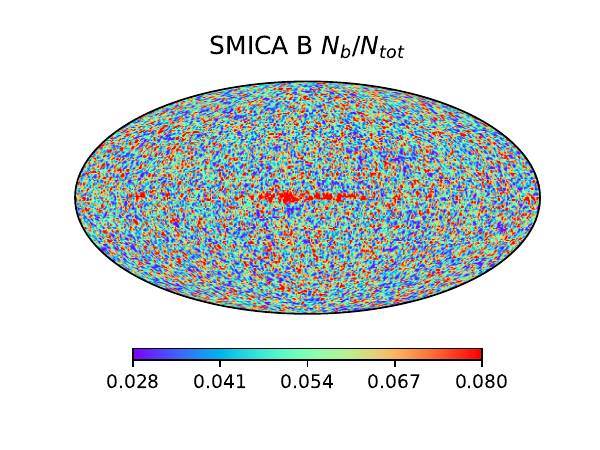}
   \includegraphics[width=0.24\textwidth]{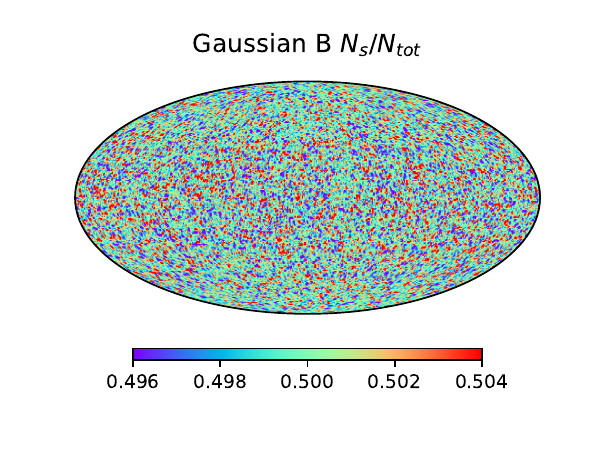}
   \includegraphics[width=0.24\textwidth]{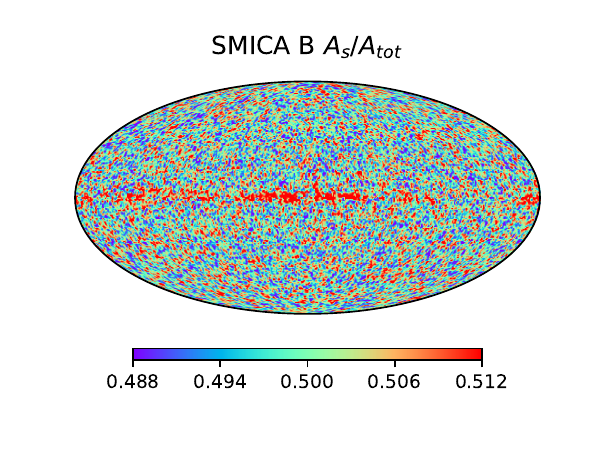}
   \includegraphics[width=0.24\textwidth]{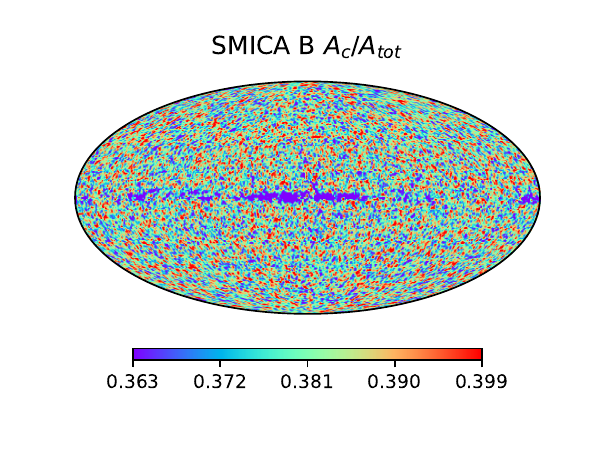}
   \includegraphics[width=0.24\textwidth]{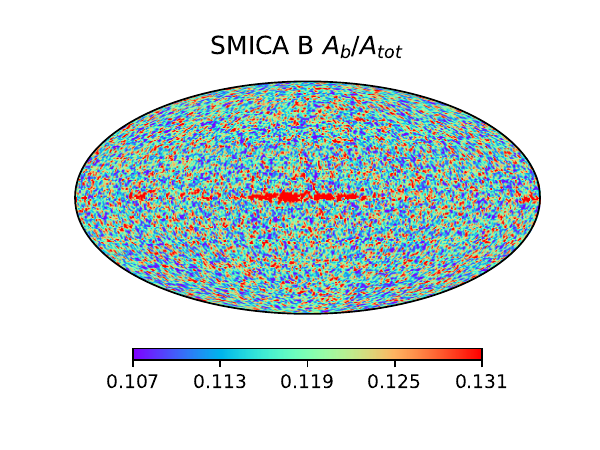}
   \includegraphics[width=0.24\textwidth]{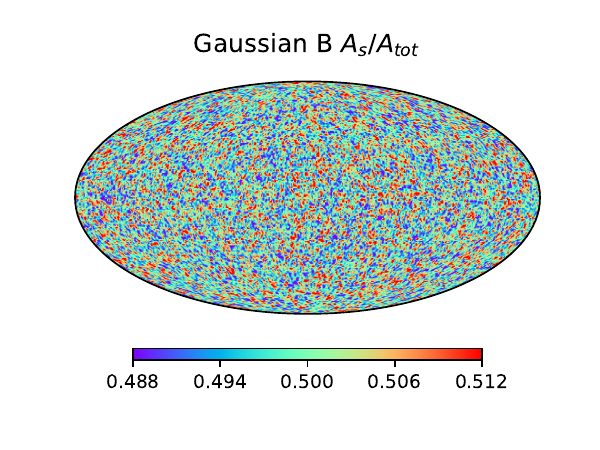}
   \includegraphics[width=0.24\textwidth]{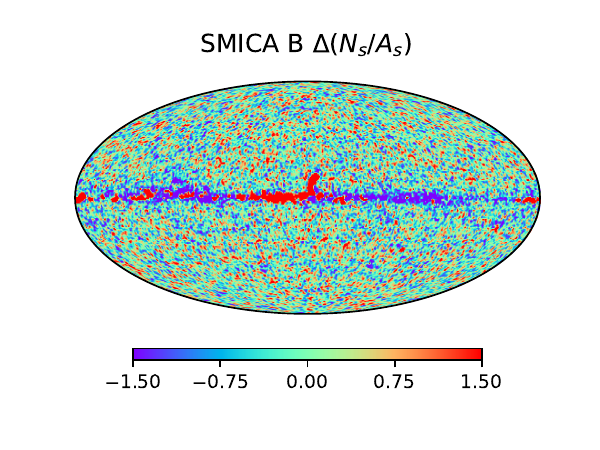}
   \includegraphics[width=0.24\textwidth]{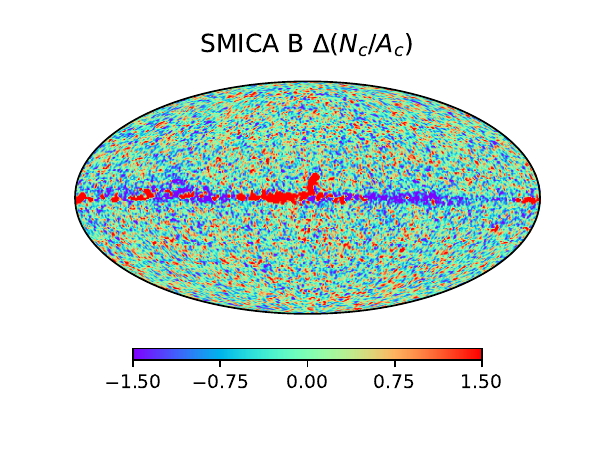}
   \includegraphics[width=0.24\textwidth]{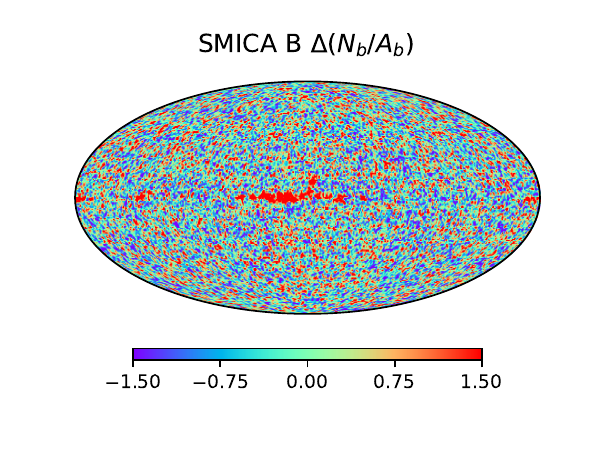}
   \includegraphics[width=0.24\textwidth]{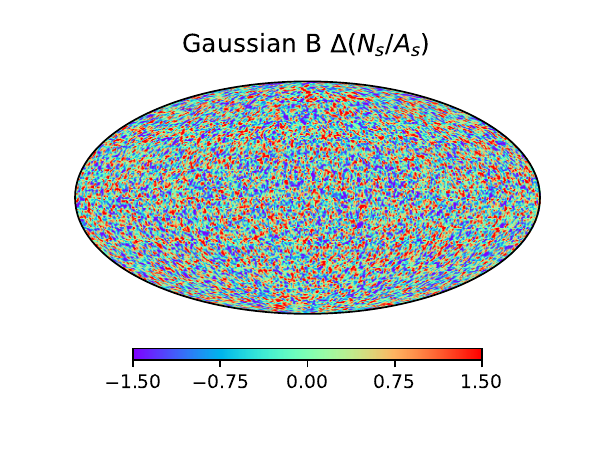}  
  \caption{Same as in Fig. 5, but for the B mode CMB SMICA Planck polarization.} 
\end{figure*}

In Fig. 4 we show the results of the ``Earth polarization" analysis for the concentration of points of
three types and their corresponding area fractions.
As it is easy to see, the concentrations of all three types of points $\tilde{n}_{type}$ deviate from the average by far more than the standard
deviation  $\langle(\tilde{n}_{type}^G)^2\rangle^{\frac{1}{2}}$ at almost all angular scales. As for the
corresponding fractions of the areas $\tilde{a}_{type}$, their deviations exceed
several thousand $\langle(\tilde{a}_{type}^G)^2\rangle^{\frac{1}{2}}$. 

To study local non-Gaussianities, it is convenient to use a map of points and areas
smoothed using the Gaussian convolution with some characteristic angular scale $\theta_0$.
Figure 4 (three lower rows) shows such maps for $\theta_0=0.5^\circ$. Here we observe an extremely nonuniform distribution of singularities
around the globe. Nonpolarized points of all types are located
mainly on the bottom of the Pacific Ocean, the Himalayas, and Australia. The areas corresponding to
saddles occupy the continents, and the areas of beaks and comets are located mainly on the
ocean shelf.

As for the distribution of the spatial concentration of points over the sphere
$\frac{N_{type}}{A_{type}}$, we show in Fig. 4 the deviation of this value from the average:
$\Delta\left(\frac{N_{type}}{A_{type}}\right)=\frac{N_{type}}{A_{type}}-\langle\left(\frac{N_{type}}{A_{type}}\right)\rangle$ (lower row).

In the same Fig. 4 we show for comparison how such maps should look for a typical implementation of the
Gaussian process with the angular spectrum of the Earth.

The anomalies in Fig. 4 demonstrate the random
yet entirely non-Gaussian nature of our planet surface formation.

\subsection{ Gaussianity test of SMICA polarization maps}

Our analysis of Planck CMB polarization data shows that the SMICA maps, although they demonstrate a strong non-Gaussianity,
do not deviate from it as radically as Commander, SEVEM and NILC. A similar result was obtained in
\citep{2025PhRvD.111f3536N} for the total number of unpolarized points of all types.
Therefore, in this article we limit ourselves to demonstrating the polarization analysis of SMICA.
We examined separately the E and B polarization maps both for the entire sky and using the conventional mask
\citep{2020A&A...641A...4P}.

\begin{center}
  \textit{E mode}
\end{center}

Figure 5 shows the results of the E polarization mode Gaussianity test at different angular scales and
the presence of the most noticeable non-Gaussian foregrounds on the polarization maps.

(1) Singularities:  $\Delta\tilde{n}_{type}(\ell_{max})$.
Significant deviations from the mean in the absence of a mask are found for comets
and beaks. With a mask, the situation improves, but the saddles deviate significantly from the expected mean
at scales corresponding to $\ell_{max}\sim 250-800$ and $\ell_{max}\sim 1700$. The same resonance scale
corresponds to beaks and comets.

It should be noted that the saddle point situation worsens after masking. This is explained
by the topology of a closed space without any holes, such as the full sphere. For most random
continuous tensor fields on the whole sphere with a large number of singularities, the number
of saddle points tends to half the total number of zero points, even if these fields are not
Gaussian. One such example is the non-Gaussian E polarization of the Earth
(see Fig. 4, upper left panel). Therefore, when calculating the total number of saddle points,
the entire sky should not be used as a test for Gaussianity. It is better to either use different
parts of the celestial sphere separately or apply a mask that eliminates this topological property.
Note that, unlike saddle points, this property does not extend to comets and beaks. The ratio
of the number of such points to the total number of singularities can be used as a test for
Gaussianity for both incomplete sky coverage and the entire celestial sphere. 

(2) Area fractions $\Delta\tilde{a}_{type}(\ell_{max})$. In the absence of a mask, huge deviations from Gaussianity arise due to foregrounds from the
Galaxy. With the mask, anomalous behavior of the functions remains at resonance scales
$\ell_{max}\sim250$ and $\ell_{max}\sim1900$.

(3) Number of points per corresponding area: $\Delta\tilde{\rho}_{type}(\ell_{max})$.
At scales $\ell_{max}>1000$, deviations from the expected Gaussian distribution exceed several
standard deviation $\langle(\tilde{\rho}_{type}^G)^2\rangle^{\frac{1}{2}}$. This is explained by the lack of a total number of unpolarized points
in the Planck maps compared to the average expected for such a spectrum for Gaussian field.

In the same figure we show maps of unpolarized points of three types and their corresponding areas, found for
the maximum possible resolution $\ell_{max}=2048$. Due to the fact that in this case the total number of such
points exceeds 2,000,000, their concentration maps were found using the Gaussian smoothing
angle of $\theta_0=0.5^\circ$. On these maps one can identify foregrounds from the Milky Way and
individual point sources.
At the same time, the ratio of the concentrations of points to the concentrations of the corresponding
areas $\Delta\frac{N_{type}}{A_{type}}$ clearly highlights not only the Galactic equator, but also the uneven
distribution
of photon noise. This is explained by the specifics of the sky scanning by the Planck mission.
Due to lower photon noise near the ecliptic poles, the observed signal is smoother
and contains fewer singularities of all types. Thus, in these regions the spectral parameters
vary locally and create a noticeable inhomogeneity in the distribution of zero polarization
points.

\begin{center}
  \textit{B mode}
\end{center}
Similar to the E mode, we analyzed the features of the observed B mode polarization.

Figure 6 for the $\Delta\tilde{n}_{type}(\ell_{max})$ and $\Delta\tilde{a}_{type}(\ell_{max})$
functions shows non-Gaussian features at approximately the same angular scales
as for the E mode. This may indicate a common origin of non-Gaussianities of characteristic
angular sizes for the E and B modes, as well as a possible E to B power leakage.
The relative concentration of saddles greatly exceeds the average expected for a Gaussian field at
almost all angular scales.
Note that, in contrast to
the results for the E mode, the function $\Delta\tilde{\rho}_{type}(\ell_{max})$  for the B mode
shows a large excess of unpolarized points of all three types. 

B mode sky maps show local non-Gaussian foregrounds in the same places where such foregrounds
manifest themselves in the E mode. However, the nonuniformity of the photon noise is not visible
on these maps. This can be explained as follows. In the E mode at least two components dominate: the
smooth cosmological signal and the photon noise. Close to the ecliptic poles the photon noise is smaller
than the average noise over the sky. Therefore, the E mode behaves more smoothly in these regions and,
as a consequence, these regions contain a smaller number of unpolarized points. For the B mode, the
cosmological signal is extremely small compared to the photon noise or is completely absent. Thus, the
local decrease in photon noise near the ecliptic poles reduces the amplitude of the Stokes parameters
in these regions, but does not change the concentration
of lines, where these parameters cross zero. As a result, such a local
change in spectral parameters does not affect the concentration of singularities.

\section{Conclusions}

One of the main tasks of modern observational
cosmology is to detect the B mode of the CMB linear polarization, which will serve as
evidence of the inflationary theory of the early Universe.
Establishing the nature of the observed B polarization is extremely important in
order to distinguish its initial cosmological component from polarized
foregrounds of cosmic and instrumental origin, as well as from distortions of
polarization maps due to gravitational lensing.

A distinctive feature of the initial scalar and tensor perturbations born at the inflation stage
is their Gaussianity.

Thus, the Gaussianity of the observed CMB polarization is a
necessary condition to ensure that the analyzed signal came to us from the surface
of the last scattering and was not caused by other radiation sources and extraneous effects.

In light of the ongoing and expected future B mode measurement experiments we need methods for
analyzing polarization maps that are most sensitive to non-Gaussian manifestations of the
two-dimensional random tensor fields.

The detailed analysis of the Planck CMB polarization maps using the statistics of singularities
(saddles, comets, and beaks) revealed non-Gaussianities in both E  and B modes at different angular scales,
which we attribute to residual foreground contamination, anisotropic noise, and possible $E\rightarrow B$ power
leakage.

The statistics of unpolarized points provide a multifaceted and highly sensitive probe of CMB polarization
Gaussianity. This method offers crucial information for diagnosing the origin of detected signals.
For the experiments like Simons Observatory \citep{2019JCAP...02..056A}, applying this technique will be vital to verify the nature of any
detected B mode signal by confirming its Gaussian character and ruling out significant foreground
contamination.  Our publicly available software can help conduct efficient and rapid Gaussianity
testing of polarization maps for experiments with incomplete or full sky coverage.

This work was supported by the RSF (Project No. 24-22-00230).

\appendix
\numberwithin{equation}{section}
\renewcommand{\theequation}{A\arabic{equation}} 
\setcounter{equation}{0}
\section*{APPENDIX}

For convenience of finding unpolarized points by the method described in  \citep{2025PhRvD.111f3536N} we
use equidistant pixelization on both spherical coordinates: $\theta_j=h\cdot j$,
$\varphi_k=h\cdot k$, $0\le j\le N$, $0\le k\le 2N$, where $h=\pi/N$. 

To simulate a map $E(\theta_j,\varphi_k)$ (or $B(\theta_j,\varphi_k)$)  whose signal consists only of harmonics with a
single fixed $\ell$, we use an
approach that allows us to create such a map in $O(N^2 \log_2N)$ operations,
\begin{equation}
  \begin{array}{l}
    E(\theta_j,\varphi_k)=\sum\limits_{m=-\ell}^\ell a_{\ell m}\tilde{P}_\ell^m(\cos\theta_j)e^{im\varphi_k},
\end{array}
\end{equation}
where the normalized associated Legendre polynomials $\tilde{P}_\ell^m(\cos\theta)$ are,
\begin{equation}
  \begin{array}{l}
    \tilde{P}_\ell^m(\cos\theta)=\sqrt{\frac{1}{4\pi}
      \frac{1}{2\ell+1}\frac{(\ell-m)!!}{(\ell+m)!!}}P_\ell^m(\cos\theta).
\end{array}
\end{equation}

Unlike the method based on successive calculation of associated Legendre polynomials with increasing $\ell$ and
fixed m, a similar approach using a recurrence formula with fixed $\ell$ and increasing or decreasing m is
unstable for small $\theta$.

Therefore, for fixed $\ell$ we use a tridiagonal system of $\ell+1$ linear equations ($0\le m\le\ell$),
where we specify boundary conditions at two ends, i.e. for $m=0$ and $m=\ell$. The system of equations looks
as follows,
\begin{equation}
  \begin{array}{l}
    \vspace{0.2cm}
    x_{\ell,j}^m\cdot\tilde{P}_{\ell,j}^{m-1}+
    y_{\ell,j}^m\cdot\tilde{P}_{\ell,j}^m+
    z_{\ell,j}^m\cdot\tilde{P}_{\ell,j}^{m+1}=0,\\
    \vspace{0.2cm}
    x_{\ell,j}^m=\sqrt{(\ell-m+1)(\ell+m)}\cdot\sin\theta_j,\\
     \vspace{0.2cm}
     y_{\ell,j}^m=2m\cdot\cos\theta_j,\\
      \vspace{0.2cm}
      z_{\ell,j}^m=\sqrt{(\ell+m-1)(\ell-m)}\cdot\sin\theta_j,\\
      x_{\ell,j}^0=0,\hspace{0.2cm}z_{\ell,j}^\ell=0,
\end{array}
\end{equation}
where we use the notation $\tilde{P}_{\ell,j}^m=\tilde{P}_\ell^m(\cos\theta_j)$.
In order to establish the boundary conditions, we must calculate the polynomials
$\tilde{P}_{\ell,j}^{0}$ and $\tilde{P}_{\ell,j}^{\ell}$  using
stable recurrence formulas,
\begin{equation}
  \begin{array}{l}
    \vspace{0.2cm}
    \tilde{P}_{0,j}^0=\frac{1}{2\sqrt{\pi}},\\
    \vspace{0.2cm}
    \tilde{P}_{n,j}^0= \alpha_n\cos\theta_j\tilde{P}_{n-1,j}^0+
    \beta_n\tilde{P}_{n-2,j}^0,\\
    \vspace{0.2cm}
    \alpha_n=\frac{\sqrt{(2n-1)(2n+1)}}{n},
    \hspace{0.1cm}\beta_n=-\frac{n-1}{n}
    \sqrt{\frac{2n+1}{2n-3}},\hspace{0.1cm}\\
\vspace{0.2cm}
\tilde{P}_{n,j}^n=-\sqrt{\frac{2n + 1}{2n}}\sin\theta_j \tilde{P}_{n-1,j}^{n-1},\\
n=1,...,\ell,\hspace{0.2cm}j=0,...,N.
\end{array}
\end{equation}

Combining Eqs. (A3), (A4) we find the coefficients for all $m$ and $j$ in $\ell\cdot N$ operations.
This method of evaluating associated Legendre polynomials with fixed boundary conditions at both ends is stable.
The remaining
part of the problem is solved using fast Fourier Transform along the $\varphi$ coordinate for
all $\theta_j$, which takes $O(N^2\log_2N)$ operations.

The standard approach to transform spherical harmonics, described in HEALPix, takes $O(N^3)$ operations.

Note that this gain in computational time occurs only if we need to
transform harmonics with a single fixed $\ell$.

The described method allows us to calculate the scalar E and pseudoscalar B fields on the sphere in a fast way.
In order to obtain the Stokes parameters and their first covariant derivatives [see Eqs. (3),(4),(6)]
we will need the derivatives of the associated Legendre polynomials over $\theta$ up to the third order.
They look as follows:
\begin{equation}
  \begin{array}{l}
    \vspace{0.2cm}
    \frac{d\tilde{P}_{\ell,j}^m}{d\theta} = \sqrt{(\ell-m)(\ell+m+1)} \cdot \tilde{P}_{\ell,j}^{m+1} + \\
\vspace{0.6cm}
m \cot\theta_j \cdot \tilde{P}_{\ell,j}^m,\\
\vspace{0.2cm}
\frac{d^2\tilde{P}_{\ell,j}^m}{d\theta^2} = [ -(\ell-m)(\ell+m+1) +\\
\vspace{0.2cm}
m^2 \cot^2\theta_j - \frac{m}{\sin^2\theta_j} ]\tilde{P}_{\ell,j}^m-\\
     \vspace{0.6cm}
    \sqrt{(\ell-m)(\ell+m+1)} \cot\theta_j \cdot \tilde{P}_{\ell,j}^{m+1},\\
     \vspace{0.2cm}
     \frac{d^3\tilde{P}_{\ell,j}^m}{d\theta^3} = \cot\theta_j[ (2m - 3m^2) \frac{1}{\sin^2\theta_j}-\\
       \vspace{0.2cm}
       (m-1) (\ell-m)(\ell+m+1) +\\
      \vspace{0.2cm}
      m^3 \cot^2\theta_j] \tilde{P}_{\ell,j}^m-\sqrt{(\ell-m)(\ell+m+1)}\times\\
    \vspace{0.2cm}
    [ (\ell-m-1)(\ell+m+2) -\\
      \vspace{0.2cm}
      (m^2 + 3m + 3) \cot^2\theta_j +(3m+1) \frac{1}{\sin^2\theta_j}]
    \tilde{P}_{\ell,j}^{m+1}.\\
\vspace{0.2cm}
    \end{array}
\end{equation}



\newpage

\def\apj{Astrophys.~J}
\def\apjl{Astrophys.~J.,~Lett}
\def\apjs{Astrophys.~J.,~Supplement}
\def\an{Astron.~Nachr}     
\def\aap{Astron.~Astrophys}
\def\mnras{Mon.~Not.~R.~Astron.~Soc}
\def\pasp{Publ.~Astron.~Soc.~Pac}
\def\aaps{Astron.~and Astrophys.,~Suppl.~Ser}
\def\apss{Astrophys.~Space.~Sci}
\def\ibvs{Inf.~Bull.~Variable~Stars}
\def\japa{J.~Astrophys.~Astron}
\def\na{New~Astron}
\def\aspproc{Proc.~ASP~conf.~ser.}
\def\aspcs{ASP~Conf.~Ser}
\def\aj{Astron.~J}
\def\actaa{Acta Astron}
\def\araa{Ann.~Rev.~Astron.~Astrophys}
\def\caosp{Contrib.~Astron.~Obs.~Skalnat{\'e}~Pleso}
\def\pasj{Publ.~Astron.~Soc.~Jpn}
\def\memsai{Mem.~Soc.~Astron.~Ital}
\def\astl{Astron.~Letters}
\def\aipproc{Proc.~AIP~conf.~ser.}
\def\physrep{Physics Reports}
\def\jcap{Journal of Cosmology and Astroparticle Physics}
\def\baas{Bulletin of the AAS}
\def\ssr{Space~Sci.~Rev.}
\def\azh{Astronomicheskii Zhurnal}

\FloatBarrier 

\bibliography{a1pol.bib}



\end{document}